\documentclass[prd,amsmath,nobibnotes,nofootinbib,superscriptaddress,preprintnumbers]{revtex4}

\usepackage{bm} \usepackage{graphicx}
\def\ba{\begin{eqnarray}}
\def\ea{\end{eqnarray}}
\def\be{\begin{equation}}
\def\ee{\end{equation}}
\def\({\left(}
\def\){\right)}
\def\[{\left[}
\def\]{\right]}
\def\<{\left<}
\def\>{\right>}
\def\cM{\mathcal{M}}
\def\cN{\mathcal{N}}

\begin{document}

\title{Runaway dilatonic domain walls}
\date{\today}

\author{Anthony Aguirre}
\email{aguirre@scipp.ucsc.edu}
\affiliation{SCIPP, University of California, Santa Cruz, CA 95064, USA}
\author{Matthew C Johnson}
\email{mjohnson@theory.caltech.edu}
\affiliation{California Institute of Technology, Pasadena, CA 91125, USA} 
\author{Magdalena Larfors}
\email{Magdalena.Larfors@physik.uni-muenchen.de}
\affiliation{Arnold Sommerfeld Center f\"ur Theoretische Physik, Ludwig Maximilians Universit\"at, M\"unchen, Germany. }

\begin{abstract}
We explore the stability of domain wall and bubble solutions in theories with compact extra dimensions. The energy density stored inside of the wall can destabilize the volume modulus of a compactification, leading to solutions containing either a timelike singularity or a region where space decompactifies, depending on the metric ansatz. We determine the structure of such solutions both analytically and using numerical simulations, and analyze how they arise in compactifications of Einstein--Maxwell theory and Type IIB string theory. The existence of instabilities has important implications for the formation of networks of topological defects and the population of vacua during eternal inflation. 
\end{abstract}

\preprint{CALT-68.276, LMU-ASC 54/09}

\maketitle

%%%%%%%%%%%%%%%%%%%

\section{Introduction}

The introduction of extra dimensions has long been one of the most useful ideas in the theorist's tool box, providing insight into important theoretical puzzles ranging from the hierarchy problem to the unification of forces and a theory of quantum gravity. However, it has been appreciated since the early days of general relativity that there will always be a proliferation of low-energy effective theories resulting from compactification of extra dimensions. This idea has gained particular force in the context of string theory, where extra dimensions are required for theoretical consistency, and the program of moduli stabilization yields a vast set of low-energy effective theories, referred to as the string theory landscape (or simply ``the landscape"). 

Classically, vacua in the landscape are stable, and drive inflation if they have positive energy density (as is the case in our universe if the observed dark energy is a cosmological constant). Quantum mechanically, they can decay via the nucleation of bubbles containing some new phase, a process described in many cases by the Coleman--de Luccia (CDL) instanton~\cite{Coleman:1980aw,Coleman:1977py}. Because the rate of bubble formation is exponentially suppressed, there is typically less than one bubble formed per Hubble volume per Hubble time, and therefore the inflating vacuum is never completely eaten up. This phenomenon is known as eternal inflation, and is a mechanism by which many different vacuum phases can be populated in different spatial regions.

Given that the postulated existence of compactified extra dimensions is the primary motivation for considering landscapes of four-dimensional vacua, it is important to determine what role extra dimensions might play in the dynamics of eternal inflation. In the four-dimensional effective theory, the overall volume of a compactification appears as a dynamical scalar field that inherits dilatonic couplings to the other low-energy fields. Therefore, the minimal model of a landscape obtained from a theory with extra dimensions contains at least two fields: one with properties determining the four-dimensional vacua (for example a scalar or gauge field), and the field related to the volume, which we will refer to as the volume modulus. 

The dilatonic coupling to the volume modulus can cause a general obstruction to finding non-singular static domain walls and CDL bubbles connecting four dimensional vacua. Because of the coupling, domain walls can perturb the volume modulus out of its vacuum, in some cases making it impossible to satisfy the appropriate boundary conditions. This was first observed by Cvetic and Soleng~\cite{Cvetic:1994ya,Cvetic:1995rp} and was applied to the landscape of Type IIB string theory by two of the authors~\cite{Johnson:2008vn}. More recently, Yang has studied these effects in 6D Einstein--Maxwell theory~\cite{Yang:2009wz}. A similar obstruction in more general multifield models can prevent the existence of other topological defects, such as monopoles~\cite{Kumar:2009pr} and strings~\cite{Achucarro:2001ii,Achucarro:1998er,Penin:1996si,Yajnik:1986tg,Yajnik:1986wq}.  

The static domain wall and CDL bubble solutions result from imposing a rather restrictive ansatz on the metric and fields, which if relaxed, may allow for additional solutions that interpolate between different four-dimensional vacua. To assess this possibility, we revisit and extend our previous work~\cite{Johnson:2008vn} to domain walls and bubbles assuming only planar, spherical, or hyperbolic symmetry. Using a toy model, we find that whenever static solutions do not exist, planar walls decay into a region where the extra dimensions decompactify. Bubble walls can also seed an expanding region where the extra dimensions decompactify. When the phase inside of the bubble has positive or zero energy density (and in some cases when the interior has negative energy density), unstable walls cause the entire bubble interior to decompactify unless the initial radius is larger than the true vacuum Hubble constant. In this case, even though the initial data contains regions of two different four-dimensional vacua, at late times the configuration contains only the original four-dimensional vacuum and a region of the decompactified phase. We apply some simple bounds to assess the stability of domain walls and bubbles in model landscapes arising from Type IIB string theory and D-dimensional Einstein--Maxwell theory, extending the analysis in Refs.~\cite{Johnson:2008vn,Yang:2009wz}.

Even when static solutions do exist, they are not always stable to perturbations. Cosmologically, this is important for the formation of networks of domain walls or other topological defects in the early universe. Another type of perturbation that is particularly relevant for eternal inflation is the collision between bubbles. We find that in many cases, decompactification can result to the future of bubble collisions, which has important implications for the observability of collisions in a realistic landscape scenario (see Ref.~\cite{Aguirre:2009ug} for a review on the observability of bubble collisions).

The instability of domain walls restricts the types of four-dimensional vacua that can be populated by eternal inflation. Although the CDL instanton cannot describe the creation of a region of some new vacuum when the bubble wall is unstable, we expect other types of transitions to mediate vacuum decay. If any such process has a classical description at late times, then even though these transitions might initially seed a region of some other four-dimensional vacuum, the unstable wall will quickly eradicate it, leaving behind a bubble containing only the decompactified phase. This might break the landscape up into sets of disconnected islands~\cite{Clifton:2007en}, with the members of each island populated only in separate  realizations of an eternally inflating universe. There may even be many eternally inflating vacua that cannot populate any other four-dimensional vacua in the landscape. These considerations make it harder to ignore the question of initial conditions, a problem that eternal inflation is often purported to eliminate. The restriction on purely four-dimensional transitions also motivates a more general and comprehensive study of the possible dynamics during the early universe in theories with extra dimensions. For example, the dynamical compactification mechanism of Ref.~\cite{Carroll:2009dn} can populate four-dimensional vacua from a higher dimensional space, leading to a multiverse where vacua of different dimensionality coexist.

The plan of the paper is as follows. In Sec.~\ref{sec:toymodel}, we introduce our two-field toy model. We review the criteria for the existence of static domain walls in the toy model, and perform a numerical analysis of non-static solutions in Sec.~\ref{sec:domainwalls}. Bubble solutions are analyzed analytically and numerically in Sec.~\ref{sec:bubbles}. We then briefly describe the implications of extra dimensions for the stability of bubble collisions in Sec.~\ref{sec:collisions}. The stability of domain walls and bubbles in the Type IIB and Einstein--Maxwell landscapes are described in Sec.~\ref{sec:EMandtypeIIB}, and we conclude in Sec.~\ref{sec:conclusions}.

%%%%%%%%%%%%%%%%%%%%%%%%%%%%%%%%%%

\section{Toy model}\label{sec:toymodel}

The compactification of extra dimensions generically gives rise to many vacuum states in the four-dimensional effective theory. The detailed properties of the vacua and the potential landscape that connects them depend on the assumptions made about the fundamental theory, and we explore several representative examples in Sec.~\ref{sec:EMandtypeIIB}. However, without going into the specifics of these constructions, we can define a toy model that encapsulates the most important generic ingredients. This will allow us to distill the relevant physics governing the stability of domain wall and bubble solutions. In particular, our toy model consists of:
\begin{itemize}
\item A field $\phi$, whose potential $V_0 (\phi)$ has multiple minima. These vacuum states are the analog of flux vacua in the Type IIB or Einstein--Maxwell landscape of Sec.~\ref{sec:EMandtypeIIB}.  
\item A field $\chi$, whose potential $V_1 (\chi)$ has a minimum at finite $\chi$, and approaches zero as $\chi \rightarrow \infty$. This field is the analog of the overall volume modulus, which determines the volume of the compactified manifold.
\item A dilatonic coupling of $\chi$ to the potential $V_0$. Couplings of this type generically occur in the four dimensional Einstein frame upon compactification.
\end{itemize}

With this set of ingredients, the potential has the structure
\begin{equation}\label{eq:potential}
V = e^{-n \chi / M_{\chi}} V_0(\phi)+ V_1(\chi), 
\end{equation}
where we choose the specific functional forms for $V_0$ and $V_1$ to be
\ba
V_0(\phi) &=& \mu_{\phi}^4  \left( \frac{\phi^2}{M_{\phi}^2} - 1\right)^2, \\
V_1(\chi) &=&  \mu_{\chi}^4 \left[- e^{-2 \chi / M_{\chi}} + a e^{-\chi / M_{\chi}} + b e^{-3 \chi / M_{\chi}} \right].
\ea

\begin{figure}
\includegraphics[scale=.8]{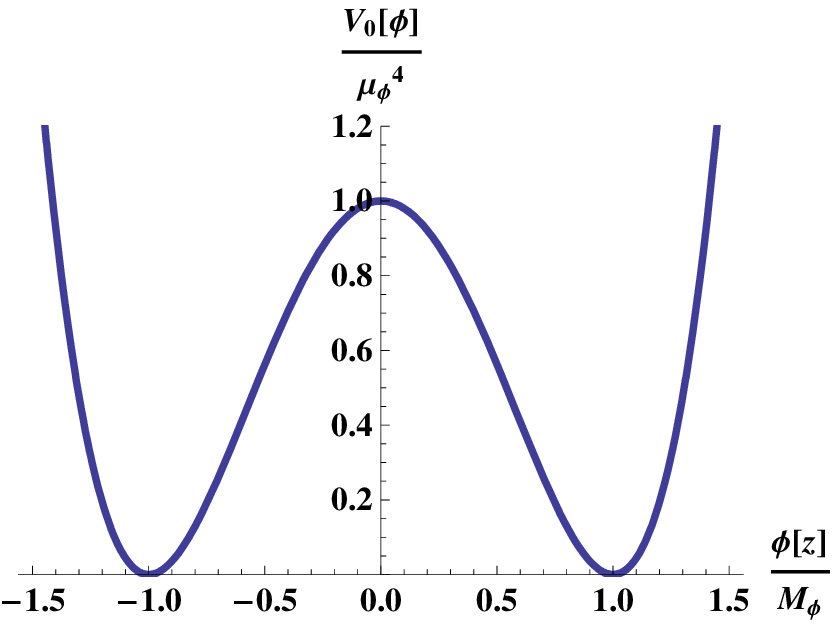}
\includegraphics[scale=.8]{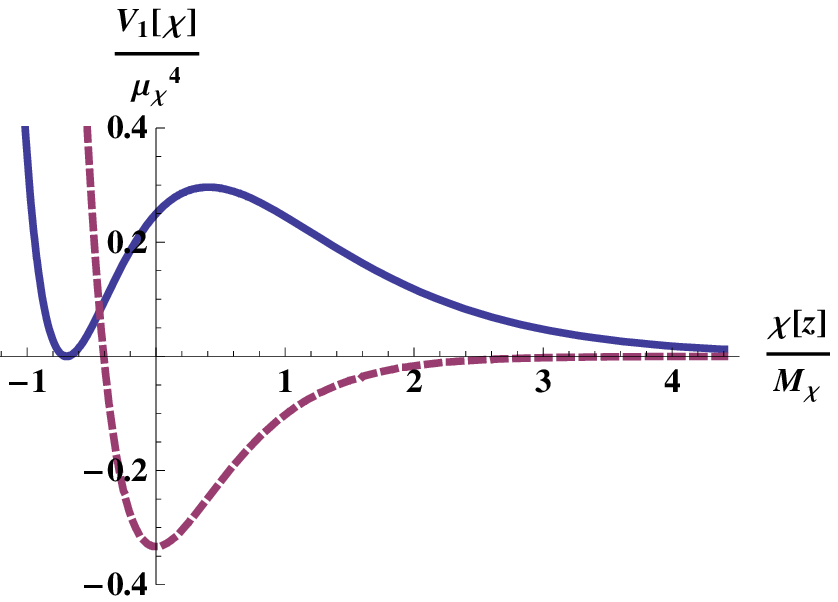}
\caption{ On the left is a plot of $V_0 (\phi)$. On the right is a plot of $V_1(\chi)$ for an AdS potential with $a=0$ (dashed) and an Uplifted potential with $b=1/4a$ (solid). 
 \label{fig:V1andV0}
}
\end{figure}

The potential $V_0 (\phi)$, shown in the left panel of Fig.~\ref{fig:V1andV0}, has two degenerate minima located at $\phi = \pm M_{\phi}$. The potential $V_1 (\chi)$ is shown in the right panel of Fig.~\ref{fig:V1andV0}. We study two limiting forms of this potential, characterized by $a=0$ and $b = 1 / 4 a$, and distinguished by their extrema and asymptotics as $\chi \rightarrow \infty$. The family of potentials with $a = 0$ resemble models with (possibly supersymmetric) Anti de Sitter (AdS) minima and the family of potentials with $b = 1 /4a$ resemble models where the potential is ``uplifted" (and supersymmetry is broken). We refer to these as ``AdS" and ``Uplifted" potentials respectively. The region of the potential where $\chi \rightarrow \infty$ corresponds to the volume of a compactification going to infinity, and we refer to this as the decompactified phase. In physical applications of the toy model, we are interested in situations where the local minima are situated at large $\chi$. Consequently, we expect $a$ to be rather small and $b$ to be rather large. 
 
For the AdS potentials, $V_1 (\chi)$ has a global minimum of depth $V_1 = -  \mu_{\chi}^4  4 / 27 b^2$ located at $ \chi_{\rm min} = M_{\chi} \log(3 b /2) $, 
and approaches zero from below as $\chi \rightarrow \infty$ (see Fig.~\ref{fig:V1andV0}, left panel). For the Uplifted potentials, $V_1 (\chi)$ has a local minimum $V_1 = 0$ at $\chi_{\rm min} = M_{\chi} \log(1 /2 a)$, a local maximum at $\chi_{\rm max} = M_{\chi} \log(3 /2 a)$ of height $V_1 = \mu_{\chi}^4 8 a^2 / 27$, and approaches zero from above as $\chi \rightarrow \infty$. 

\begin{figure}
\includegraphics[scale=.8]{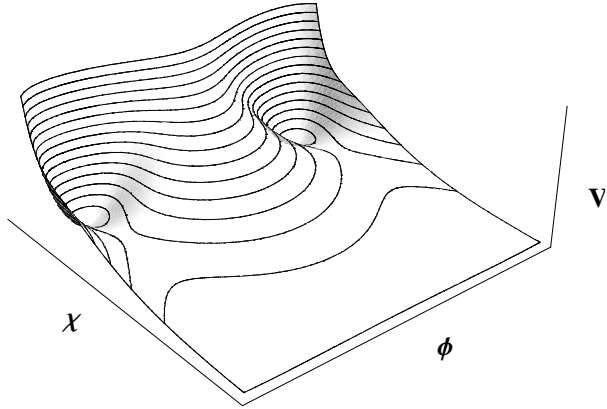}
\caption{ A sketch of the full potential in Eq.~\ref{eq:potential} for an Uplifted $V_1$. There are two local minima, between which a  runaway direction to the decompactified phase at $\chi \rightarrow \infty$ opens up.
 \label{fig:potential}
}
\end{figure}

The full potential is shown in Fig.~\ref{fig:potential} for an Uplifted $V_1$. There are two full minima located at $\{ \chi = \chi_{\rm min}, \phi = \pm M_{\phi} \}$. Depending on the parameters, it is possible that  the minima and maxima for $\chi$ can disappear at $\phi \neq M_{\phi}$, opening up a runaway direction towards large $\chi$ (as in Fig.~\ref{fig:potential}). For example, with $n=1$, this occurs for $3 b < \mu_{\chi}^4/\mu_{\phi}^4$ (with $a=0$) and $a < 3 \mu_{\phi}^4 / \mu_{\chi}^4$ (with $b = 1 / 4a$). The existence of this runaway direction will have dramatic implications for the stability of domain walls, as we now show.

\section{Stability analysis of domain wall solutions}\label{sec:domainwalls}

\subsection{Analytic results}\label{sec:DWanalytic}

Our goal in this section will be to find static solutions where a stable planar domain wall interpolates between minima of the potential Eq.~\ref{eq:potential}. Neglecting gravitational effects, and assuming planar symmetry, the fields evolve according to
\begin{equation}
\label{eq-planarevolve}
\partial_z^2 \phi = \frac{dV}{d\phi},\ \ 
\partial_z^2 \chi = \frac{dV}{d\chi},
\end{equation}
where $z$ is the coordinate perpendicular to the domain wall. One class of domain walls connects minima of the potential $V_1(\chi_{\rm min})$ to $\chi \rightarrow \infty$ with $\phi$ fixed in vacuum. Another class of domain walls interpolates between $\phi = + M_{\phi}$ and $\phi = - M_{\phi}$ with $\chi = \chi_{\rm min}$ on either side. For this latter class of solutions, because of the coupling between $\phi$ and $\chi$ in the potential Eq.~\ref{eq:potential}, both fields have a non-trivial profile inside of the wall. 

Domain walls of the first class, which involve only the $\chi$ sector, are easy to construct. The appropriate boundary conditions for the equation of motion Eq.~\ref{eq-planarevolve} are $\chi( -\infty) = \chi_{\rm min}$ and $\partial_z \chi ( -\infty) = 0$ with $\phi$ in vacuum. Evolving in an AdS potential, the field reaches $\chi( + \infty) \rightarrow \infty$ with non-zero $\partial_z \chi$.\footnote{Physical examples of these types of domain walls are often supersymmetric, and thereby stable (see e.g. \cite{Ceresole:2006iq}).}

\begin{figure}
\includegraphics[scale=.8]{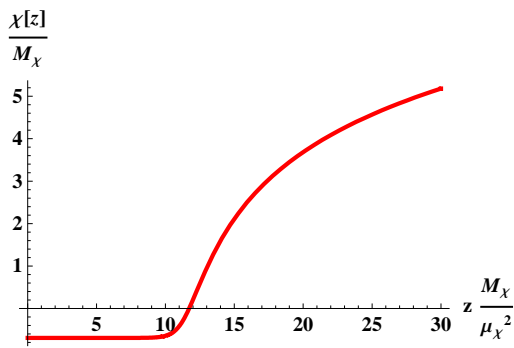}
\caption{A portion of the domain wall solution interpolating between $\chi = \chi_{\rm min}$ and $\chi \rightarrow \infty$ along one of the minima for $\phi$.
 \label{fig:largechidomainwall}
}
\end{figure}

For an Uplifted potential, the field reaches $\chi( + \infty) \rightarrow \infty$ with $\partial_z \chi (+ \infty) = 0$. A portion of such a solution is shown in Fig.~\ref{fig:largechidomainwall}. The field interpolates between $\chi_{\rm min}$ at $z \rightarrow -\infty$ and $\chi \rightarrow \infty$ at $z \rightarrow + \infty$. The tension of such walls can be calculated exactly \cite{Coleman:1980aw}:
\begin{equation}
\sigma_{\chi} = \int_{\chi_{\rm min}}^{\infty} \sqrt{2 |V_1|} d\chi = \frac{8a}{3} \mu_{\chi}^2 M_{\chi}.
\end{equation}

Domain wall solutions that interpolate between the vacua for $\phi$ have a more complicated structure. We investigate these solutions analytically by first finding domain walls between the vacua of $V_0$ at fixed $\chi = \chi^*$, and then studying their back-reaction on the $\chi$ sector. The static domain wall solution for fixed $\chi = \chi^*$ is given by:
\begin{equation}\label{eq:phidomainwall}
\phi_{DW} (z) = M_{\phi} \tanh \left[ \frac{\sqrt{2} \mu_{\phi}^2}{M_{\phi}} e^{-n \chi^* / 2 M_{\chi}} ( z - z^*) \right],
\end{equation}
where $z^*$ is the position of the domain wall. The thickness of the wall is approximately
\begin{equation}\label{eq:deltaz}
\Delta z =  \frac{M_{\phi}}{\sqrt{2} \mu_{\phi}^2} e^{n \chi^* / 2 M_{\chi}}.
\end{equation}

The tension is defined by \cite{Coleman:1980aw}
\be
\sigma_{\phi} = 2 \int_{z^* - \hat{d}}^{z^* + \hat{d}} \left[ V(\phi_{DW}, \chi^*) - V(M_{\phi}, \chi_{\rm min}) \right] dz.
\ee
To capture all of the energy density associated with the wall, we must strictly take the limit where $ \hat{d} \rightarrow \infty$. In the following, we neglect the small contribution to the tension that generically arises from the integral over $V_1$. This is consistent as long as $V_1 (\chi^*) \ll V_0$, which is generally a good approximation when $\chi^* \sim \chi_{\rm min}$ (stable domain walls) or when $\chi^* \gg \chi_{\rm max}$ (unstable domain walls).
   In order to keep the dependence on $\chi^*$ in the tension explicit, we also define $u = e^{-n\chi^*/2 M_{\chi}} z$ so that
\be\label{eq:walltension}
\sigma_{\phi} = 2 e^{-n\chi^*/2 M_{\chi}} \int_{u^* - d}^{u^* + d}  V_0 du \equiv  e^{-n\chi^*/2 M_{\chi}} \sigma_0,
\ee
where 
\begin{equation}
\sigma_0 = \frac{4 \sqrt{2}}{3} \mu_{\phi}^2 M_{\phi}
\end{equation}
 would be the tension of the domain wall if the potential for $\phi$ were simply $V (\phi) = V_0 (\phi)$. 

\begin{figure}
\includegraphics[scale=.8]{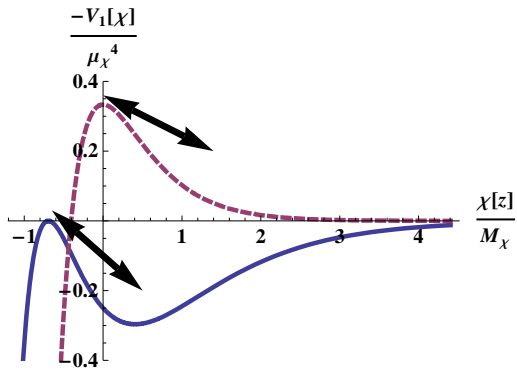}
\caption{The Euclidean potential $-V_1$ in the Uplifted (solid) and AdS (dashed) case. Domain wall solutions must interpolate to and from the Euclidean maxima of the potential.
 \label{fig:euclideanpotential}
}
\end{figure}

We now estimate how the $\chi$-field evolves in this thin domain wall background. For static solutions ($\partial_t \chi=0$), the problem reduces to determining the motion of a particle in the inverted potential $-V_1$ of Fig.~\ref{fig:euclideanpotential}, which at 'time' $z = z^*$ experiences an additional force from the domain wall. Substituting the domain wall solution into the equations of motion we obtain
\begin{equation}\label{eq:eomforchi}
\partial_z^2 \chi = - \frac{n}{M_{\chi}} e^{-n \chi / M_{\chi}} V_0 [ \phi_{DW} (z, \chi^*) ] +  \frac{d V_1}{d \chi}.
\end{equation}
Note that even when the $\chi$ field starts in its vacuum (where $d V_1 / d \chi = 0$), the domain wall necessarily induces a spatial gradient. 

When the thickness of the wall is very small compared to the other scales in the equation of motion, the domain wall for $\phi$ acts as a $\delta$-function source for the $\chi$ field. The influence of the $dV_1 / d \chi$ term in Eq.~\ref{eq:eomforchi} on the motion of $\chi$ is then negligible in the vicinity of the domain wall. Thus, we estimate the discontinuity in the spatial gradient of the $\chi$ field by integrating the equation of motion Eq.~\ref{eq:eomforchi} across the location of the domain wall. Doing so, and using the tension defined in Eq.~\ref{eq:walltension}, gives
\begin{equation}\label{eq:kicksize}
\Delta (\partial_z \chi ) \simeq - \frac{n \sigma_0}{2 M_{\chi}} e^{-n \chi^* / 2 M_{\chi}}.
\end{equation}
This relation first appeared in Ref.~\cite{Cvetic:1994ya}. 

We now return to the full equation of motion for $\chi$, including the influence of $V_1 (\chi)$. We would like to find a solution where $\chi = \chi_{\rm min}$ on either side of the domain wall. In the mechanical analog, this means that the particle must start out at the maximum of the Euclidean potential $-V_1$ at $z \rightarrow -\infty$ and end up back at the maximum as $z \rightarrow +\infty$. This is illustrated in Fig.~\ref{fig:euclideanpotential}. Because $\chi$ receives a kick at the location of the domain wall in $\phi$, staying at the maximum for all $z$ is not a possible solution. Instead, $\chi$ must gain enough velocity in the positive-$\chi$ direction to exactly balance the gain in velocity in the negative-$\chi$ direction due to the kick. A solution exists if the size of the kick can be tuned so that it matches twice the magnitude of the rolling velocity of the particle. In this case, the trajectory is time-reversed, and the particle ends up at rest at the maximum, without over- or undershooting it.

Since energy is conserved, the rolling velocity in the mechanical analog is simply given by $\partial_z \chi = \sqrt{2 |V_1 (\chi_{\rm min}) - V_1 (\chi^*)|}$, which we must compare at different values of $z$ (or equivalently, different values of $\chi^*$) to the kick Eq.~\ref{eq:kicksize}. There is a solution when, for some $\chi^*$, 
\begin{equation}\label{eq:kickcompare}
2 \sqrt{2 |V_1 (\chi_{\rm min}) -  V_1 (\chi^*)|} = \frac{n \sigma_0}{2 M_{\chi}} e^{-n \chi^* / 2 M_{\chi}}.
\end{equation}

For the AdS potentials, it is always possible to satisfy Eq.~\ref{eq:kickcompare}. At large $\chi$, the potential goes to zero, and the rolling velocity is therefore approximately given by the height of the Euclidean maximum. In the limit of large $b$, we must go out to large $\chi^*$ to satisfy Eq.~\ref{eq:kickcompare}, which, using the asymptotic rolling velocity, reduces to
\begin{equation}
\frac{2 \sqrt{2} \mu_{\chi}^2 }{3 \sqrt{3} b} = \frac{n \sigma_0}{M_{\chi}} e^{-n \chi^* / 2 M_{\chi}}.
\end{equation}
This is satisfied for
\begin{equation}\label{eq:azerochi}
\chi^* = \frac{2 M_{\chi}}{n} \log \left[ \frac{\sqrt{3} b n}{2} \left( \frac{\mu_{\phi}^2 M_{\phi} }{\mu_{\chi}^2 M_{\chi} } \right) \right].
\end{equation}
In physical applications, it is well known that BPS domain walls exist between supersymmetric vacua \cite{Cvetic:1993xe,Cvetic:1996vr,Cvetic:1995rp,Cvetic:1992st,Behrndt:2001mx,LopesCardoso:2001rt,LopesCardoso:2002ec,Behrndt:2002ee,Ceresole:2001wi,Louis:2006wq, Ceresole:2006iq}. The domain walls we have constructed in our toy model are analogous these types of solutions. 
  
The story is very different for the Uplifted potentials. In this case, the Euclidean potential $-V_1$ has a local minimum, past which the rolling velocity decreases as $\chi \rightarrow \infty$. Consequently, it is possible to find situations where the kick always overwhelms the motion in the potential. If this occurs, the static domain wall solution does not exist. Just as for the AdS potentials, a static solution only exists when Eq.~\ref{eq:kickcompare} can be satisfied for some $\chi^*$. Without using the full potential $V_{1} (\chi)$, there are two simple tests we can apply to check this condition. These will be convenient in our analysis of physical models in Sec.~\ref{sec:EMandtypeIIB}, where the potential for $\chi$ takes a complicated form.

For the first test, we can compare the rolling velocity at the minimum of the Euclidean potential (where it is maximized) to the kick velocity. In this case, when the inequality
\begin{equation}\label{eq:maximumcondition}
\frac{n \sigma_0}{2 M_{\chi}} e^{-n \chi_{\rm max} / 2 M_{\chi}} < 2 \sqrt{2 |V_1 (\chi_{\rm max})|},
\end{equation}
is satisfied, there exists a solution where $\chi$ does not roll past the position of the Euclidean minimum. Substituting with $\chi_{\rm max}$ and $V_1 (\chi_{\rm max})$, our toy model admits a static domain wall when
\begin{equation}\label{eq:toymodelcondition}
2^{(n-3)/2} 3^{(n-1)/2} n  \left( \frac{\mu_{\phi}^2 M_{\phi} }{\mu_{\chi}^2 M_{\chi} } \right) < a^{(2-n)/2}.
\end{equation}
This translates into a lower bound on $a$. For example, with $n = 0.5$, and $\mu_{\phi}^2 M_{\phi} = \mu_{\chi}^2 M_{\chi} $, the critical value of $a$ below which the condition for a stable domain wall solution is no longer satisfied is $a_{\rm crit} \simeq 0.18$. This lower bound on $a$ is an important restriction, since we should recall that $a$ has to be small in order for the potential to have a minimum at large $\chi$ (as is required in physical realizations of this toy model).

Re-arranging Eq.~\ref{eq:maximumcondition}, and noting that $\sigma_{\chi} \sim \sqrt{|V_1 (\chi_{\rm max})|} M_{\chi}$, we see that the condition for a solution roughly corresponds to requiring that
\begin{equation}
\sigma_{\phi} < \sigma_{\chi}.
\end{equation}
Consequently, stable domain wall solutions exist only if the tension of a $\phi$ domain wall, separating the two vacua at finite $\chi$, is smaller than the tension of a $\chi$ domain wall, separating the vacua at finite $\chi$ from the decompactified phase. If the $\phi$ domain wall has the larger tension, it is energetically favourable to form a lower-tension $\chi$ domain wall, i.e. for the solution to decompactify. We will see in the numerical results of the next section that this is indeed the case.

We now turn to the second test for domain wall stability. If the field rolls past the Euclidean minimum, it may still be possible to find a solution at large $\chi^*$. In this limit, the potential for $\chi$ can be approximated by
\begin{equation}
V_1 \simeq \mu_{\chi}^4 a e^{-\chi / M_{\chi}}. 
\end{equation}
If the asymptotic falloff of the kick velocity is faster than that of the rolling velocity, then it is always possible to find a solution at sufficiently large $\chi^*$. Comparing $\sqrt{|V_1|}$ to Eq.~\ref{eq:kicksize}, we see that this occurs for $n>1$. However, when $n \leq 1$, then if there is no solution at small $\chi^*$, there is no solution in the asymptotic region where $\chi^* \rightarrow \infty$. The parameter bounds for planar domain walls are summarized in Table~\ref{tab:bounds}.

\begin{table*}
\begin{tabular}{l l l}
\hline
\hline
AdS potential ($a=0$) & & Stable\\
Uplifted potential,  $n>1$ & &Stable\\
Uplifted potential,  $n \leq 1, a>a_{\rm crit} $ && Stable\\
Uplifted potential,  $n \leq 1, a \leq a_{\rm crit} $ && Unstable\\
\hline
\hline
 \end{tabular}
 \begin{center}
 \caption{For the AdS potential, planar domain walls are always stable to decompactification. For the Uplifted potential, stable planar domain walls are only possible for restricted values of the parameters. Recall that small $a$ is required for the potential to have minima at large values of $\chi$.
   \label{tab:bounds}}
 \end{center}
\end{table*}

When a stable solution cannot be found, the kick causes the $\chi$ field to overshoot its minimum and roll off to $\chi \rightarrow - \infty$. At large negative $\chi$, the potential can be approximated as
\begin{equation}
V_1 \simeq \frac{\mu_{\chi}^4}{4a}  e^{- 3 \chi / M_{\chi}}. 
\end{equation}
There is an attractor solution to the equations of motion for such a potential, given by
\begin{equation}
\chi (z) = - \frac{2 M_{\chi}}{3} \log \left[ \sqrt{\frac{3}{8a}} (z_0 - z) \right]. 
\end{equation}
From this solution, it can be seen that $\chi \rightarrow -\infty$ at finite $z = z_0$, and the potential energy diverges. 

The solutions we have constructed neglect the gravitational back-reaction of the walls. Perhaps the most dramatic effect of including gravity is the appearance of a planar, timelike singularity in the unstable domain wall solutions, occurring at the value of $z= z_0$ where the potential diverges. Even for stable walls, the spacetime is altered in important ways. Due to the repulsive gravitational field of domain walls, the spacetime cannot be static~\cite{Vilenkin:1984hy,Ipser:1983db}. Observers on both sides of the domain wall perceive themselves to be surrounded by a spherical membrane that accelerates towards them, reaches a turning point, and then accelerates away (see e.g.~\cite{Bousso:1998vz}). However, gravitational effects notwithstanding, the field dynamics we have studied above should be a good approximation to the full solution in the limit where $\sigma \ll M_4^3$ and the thickness of the wall is much smaller than $M_4^2 / \sigma$, where $M_4$ is the four-dimensional Planck mass. Our criteria for the existence of domain wall solutions are therefore widely applicable. 

\subsection{Numerical results}\label{sec:DWnumeric}

In the previous section, we identified situations where it is impossible to find a static domain wall solution that interpolates between $\phi = \pm M_{\phi}$. The requirement that the domain wall configurations are static is fairly restrictive, and it is possible that there exist stable time-dependent interpolations. The relevant equations of motion in the absence of gravity are given by
\begin{equation}\label{eq-planarevolve2}
\partial_t^2 \phi - \partial_z^2 \phi = - \frac{dV}{d\phi},\ \ 
\partial_t^2 \chi - \partial_z^2 \chi = - \frac{dV}{d\chi}.
\end{equation}
 
Without the static ansatz, it is in general no longer possible to find analytic solutions, and we employ a set of numerical simulations. These simulations should confirm the existence of static solutions over a range of parameter space, and inform us about what happens to domain wall configurations when our analytic approach predicts that static solutions cannot exist. We will see that in these cases, the unstable walls seed a growing region of the decompactified phase.

To find static solutions, we first choose one of our toy model potentials. We start with an initial guess for the field configuration $\phi (z, t=0)$ given by Eq.~\ref{eq:phidomainwall} and $\chi (z,t=0) = \chi_{\rm min}$. Adding an artificial friction term to the equations of motion, this initial data will evolve to the true static solution (if it exists) after a sufficient amount of simulation time elapses. We then check that the solution is truly static by inputting the final data from the run with friction as initial data in a run without friction, and seeing if there is any time-evolution.

\begin{figure}
\includegraphics[width=15cm]{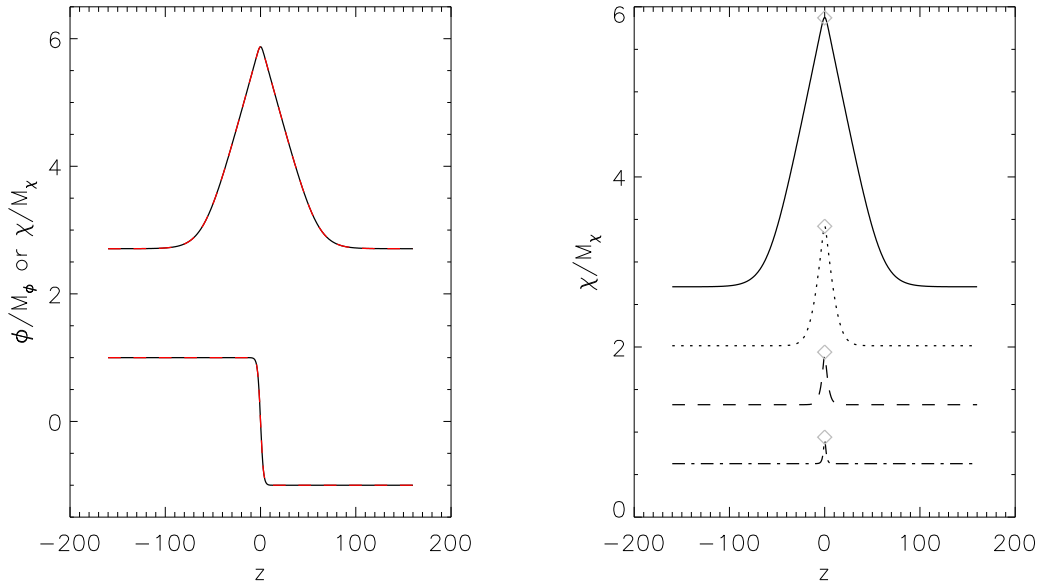}
\caption{On the left is the static domain wall configuration for an AdS potential with $b=10$. The fields at the beginning (solid black) and end (dashed red) of the simulation without artificial friction match nearly perfectly, indicating that the configuration is truly static. On the right, we show the numerically generated profile for $\chi$, with $b = \{ 1.25, 2.5, 5, 10\}$ from bottom (dot dashed) to top (solid). As $b$ is increased, the excursion of $\chi$ from its minimum increases. The prediction for the value of $\chi = \chi^*$ at the location of the domain wall (Eq.~\ref{eq:kickcompare}) is indicated by the diamond in each case. The asymptotic expression for $\chi^*$ given by Eq.~\ref{eq:azerochi} becomes accurate for $b \agt 10$.
 \label{fig:staticnumeric1}
}
\end{figure}

Applying this procedure to the AdS potentials, we are always able to find static solutions. An example for $b=10$ is shown in the left panel of Fig.~\ref{fig:staticnumeric1}, where it can be seen that the initial and final configurations in the run without the artificial friction term are identical. The form of the solution is very similar to what is predicted by the mechanical analog of the previous section: the $\chi$ field rolls out some distance, and then is reversed by the kick experienced at the location of the wall. We can perform a more detailed check of our analytic estimates by comparing the predicted amplitude of $\chi$ inside the wall, given by solving Eq.~\ref{eq:kickcompare} (or Eq.~\ref{eq:azerochi} in the limit of large $b$) for $\chi^*$, to the maximum excursion of $\chi$ seen in the simulations. In the right panel of Fig.~\ref{fig:staticnumeric1} it can be seen that there is excellent agreement over a wide range in the parameter $b$.

\begin{figure}
\includegraphics[width=15cm]{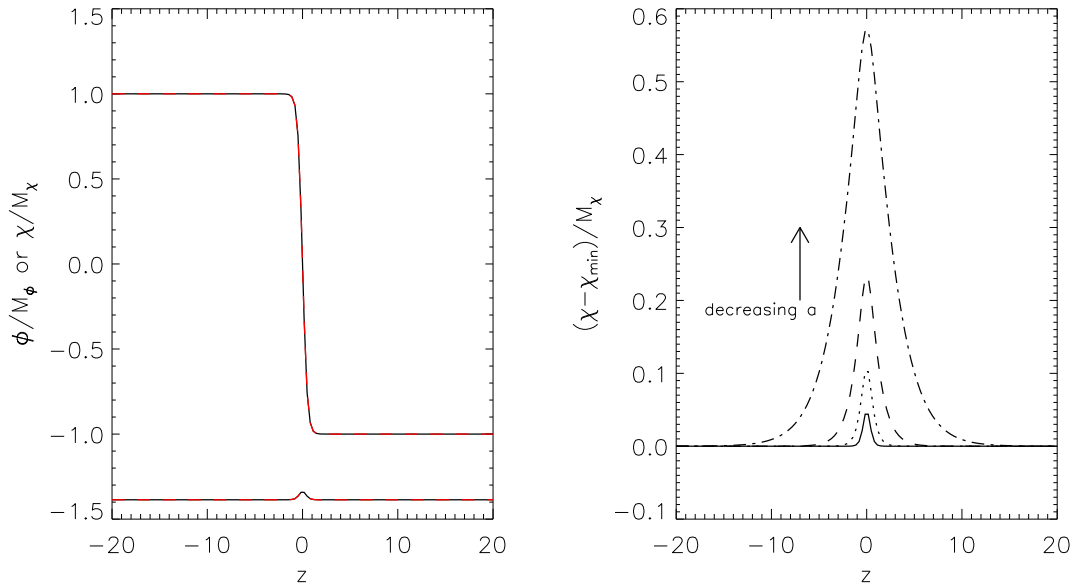}
\caption{On the left, we plot the static domain wall configuration for an Uplifted potential with $a=2$ and $n=.5$. The fields at the beginning (solid black) and end (dashed red) of the simulation without artificial friction match nearly perfectly, indicating that the configuration is truly static. On the right, we plot the static profile $\chi(z) - \chi_{\rm min}$, for $a = \{ 2,1,.5,.25\}$. Just as predicted by the analytic model, the amplitude of $\chi$ inside of the $\phi$ wall increases as $a$ decreases.
 \label{fig:chiseries}
}
\end{figure}

We now turn to the Uplifted class of potentials. In the previous section, we estimated that static solutions would only be possible when the inequality Eq.~\ref{eq:toymodelcondition} is satisfied. For fixed $n$, this yields a critical value $a= a_{\rm crit}$, below which a static solution should no longer be possible. Beginning at a value of $a > a_{\rm crit}$, we find the improved static solution by the method described above. An example of such a static solution is shown in the left panel of Fig.~\ref{fig:chiseries}. We then incrementally decrease $a$, at each step feeding in the improved static configuration from the previous step as the initial condition. In the right panel of Fig.~\ref{fig:chiseries}, we plot the static profile for $\chi(z)$. As $a$ is decreased, the amplitude of $\chi^*$ inside of the wall increases, as predicted by the analysis of the previous section.

\begin{figure}
\includegraphics[width=15cm]{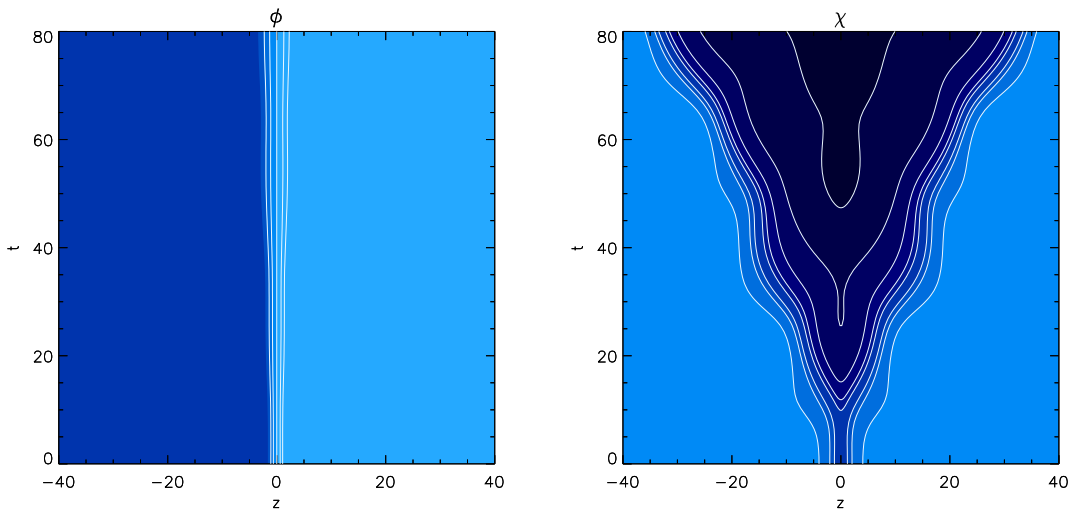}
\caption{An unstable domain wall, just below the critical value of $a$ ($a = 0.125$) with $n=.5$. On the left is a contour plot of $\phi(z,t)$ and on the right we plot $\chi (z,t)$. The initial configuration at $t=0$ is set by the static configuration for $a=.25$. Starting from the location of the $\phi$ wall, a region where the $\chi$ field rolls to $\chi \rightarrow \infty$ develops (the black region in the right panel), eventually spreading over the entire simulated region. The contours in the plot of $\chi(z,t)$ extend from $\chi_{\rm min} \simeq 1.4$ out to $\chi \simeq 4.8$, which is well past $\chi_{\rm max}$, and in the run-away region of the potential.  
 \label{fig:unstablewall}
}
\end{figure}

Continuing to decrease $a$, eventually the initial conditions from the previous step cease to evolve to a static configuration. This holds independently of the step size in $a$. Instead, as shown in Fig.~\ref{fig:unstablewall}, field configurations interpolating between the vacua of $\phi$ cause $\chi$ to roll off to $\chi \rightarrow + \infty$. If we set $n=.5$, the analytical comparison of the kick velocity to the barrier height (Eq.~\ref{eq:toymodelcondition}) predicts that the critical value of $a$ is $a_{\rm crit} \simeq 0.18$. Performing a more careful analytic analysis, in which the rolling velocity is compared to the kick velocity at all $\chi^*$ (Eq.~\ref{eq:kickcompare}), we predict $a_{\rm crit} \simeq 0.168$. Numerically, we find the critical value to be between $0.171 < a_{\rm crit} < 0.177$. The analytic estimates are in very good agreement with the numerical results, and it can be seen that the simple estimate Eq.~\ref{eq:toymodelcondition} gives a good indication of the value of $a_{\rm crit}$. We have repeated this analysis for various $n$, with similar results. For $n=.25$ we predict $a_{\rm crit} \simeq 0.11$ (from Eq.~\ref{eq:kickcompare}), $a_{\rm crit} = 0.11$ (from Eq.~\ref{eq:toymodelcondition}), and find $0.1148 < a_{\rm crit} < 0.1189$ numerically; for $n=.75$ we predict $a_{\rm crit} \simeq 0.18$ (from Eq.~\ref{eq:kickcompare}) and $a_{\rm crit} = 0.22$ (from Eq.~\ref{eq:toymodelcondition}), compared with $ 0.1895 < a_{\rm crit} < 0.1962$ from the numerics. In summary, whenever a static solution cannot be found in the Uplifted potentials, planar domain walls interpolating between the vacua at $\phi = \pm M_{\phi}$ inevitably seed decompactification.

The analytic estimates in the previous section were made for an infinitely thin wall. Numerically, we can determine how varying the thickness of the $\phi$ wall at fixed tension $\sigma_0$ affects the profile of the $\chi$-field. To do this, it is convenient to define a new parameter $\alpha = \mu_{\phi}/M_{\phi}$. At fixed $\sigma_0$, increasing $\alpha$ will decrease the thickness of the wall:
\begin{equation}
\Delta z \propto \left( \frac{2}{3 \alpha^4 \sigma_0} \right)^{1/3}. 
\end{equation}
In the limit where $\alpha \rightarrow \infty$, the numerics should reproduce the analytic results of the previous section. 

\begin{figure}
\includegraphics[width=11cm]{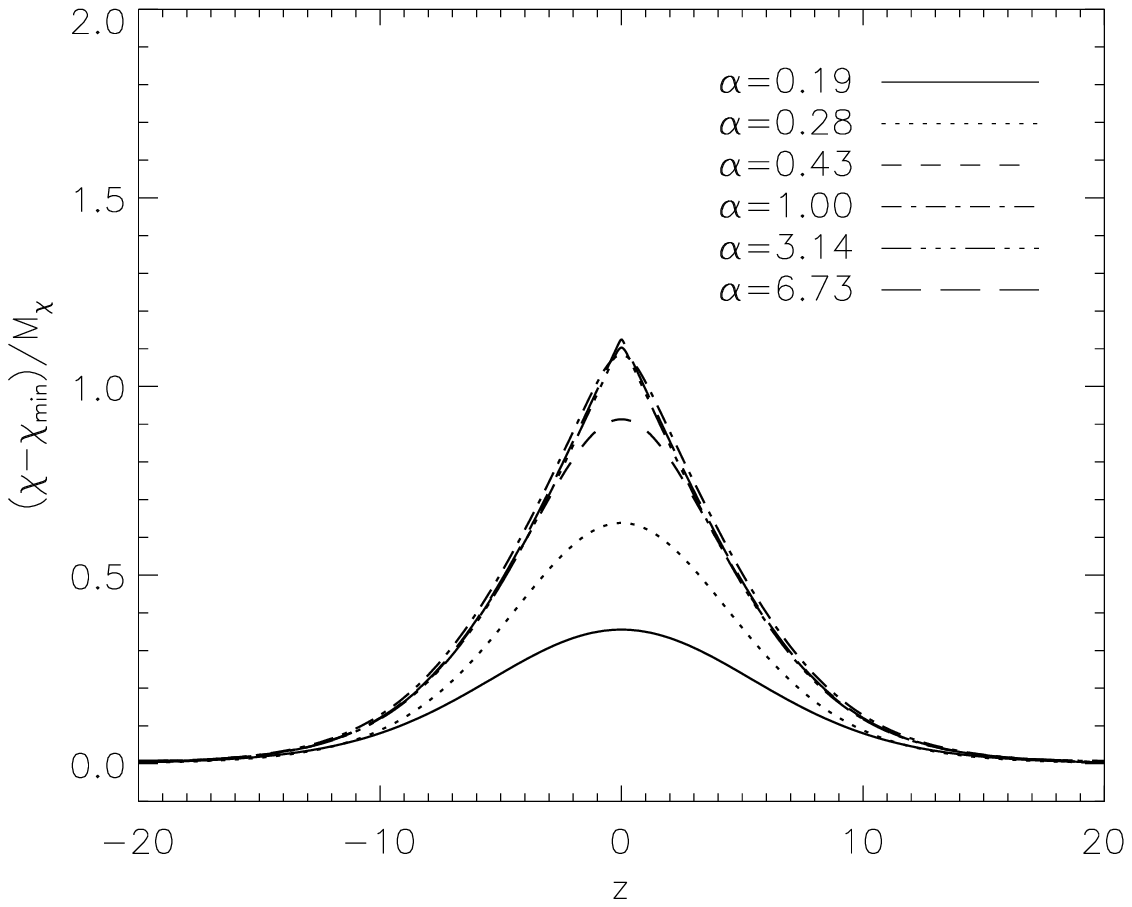}
\caption{A plot of $\chi(z)$ for static domain walls with $\{ a=.18, n=.5\}$ and varying $\alpha = \{ 0.19,.28, 0.43, 1., 3.14,6.73\}$. Increasing $\alpha$ corresponds to the $\phi$ wall becoming thinner while retaining a fixed tension $\sigma_0$. The profile $\chi (z)$ approaches a fixed shape in the limit of an infinitely thin wall, which is well reproduced by assuming a $\delta$-function source in the equation of motion for $\chi$, as in the analytic treatment of Sec.~\ref{sec:domainwalls}. The predicted maximum excursion of $\chi$ (for the above set of parameters) is $(\chi^* - \chi_{\rm min})/M_{\chi} \simeq 1.11$
 \label{fig:alphas}
}
\end{figure}

In Fig.~\ref{fig:alphas}, we show the profile of the $\chi$-field for increasing values of $\alpha$. As $\alpha$ is increased, the $\chi$ profile becomes more cuspy, and asymptotically approaches a fixed maximum excursion, $\chi = \chi^*$, inside of the $\phi$ wall. The $\phi$ wall acts as a $\delta$-function source for the $\chi$ field in this limit, in accord with the cuspy profile. Note also that the $\chi$ field becomes increasingly displaced from its vacuum for thinner walls. Quantitatively, the predicted value of $\chi^*$ for an infinitely thin wall (obtained by comparing the rolling velocity to the size of the kick in the mechanical analogue, given by Eq.~\ref{eq:kickcompare}) agrees well with the numerics: we predict $(\chi^* - \chi_{\rm min})/M_{\chi} \simeq 1.11$ compared with $(\chi^* - \chi_{\rm min})/M_{\chi} \simeq 1.15$ obtained from the numerical results.

Even in the range of $a$ where static solutions are possible, especially near $a_{\rm crit}$, the ability of the configuration to relax into a solution that separates the vacua of $\phi$ with $\chi$ remaining finite is very much dependent on the initial conditions.  As an example, if we were to set $\chi$ everywhere equal to $\chi_{\rm min}$, the value of $a$ above which $\chi$ does not roll off to infinity is reduced to $a \simeq 0.23$ from the boundary for stable domain walls at $a_{\rm crit} \simeq 0.17$. For a more complicated initial configuration, where the field is sprinkled randomly over different regions of the potential, the stability of various domains to decompactification is undoubtedly more pronounced. This ``over-shoot problem" is certainly relevant for cosmological mechanisms of domain wall production. As an example, one can consider the Kibble mechanism, where different spatial regions of an initially hot universe relax into different vacua as the universe cools down. Without fine-tuning the initial conditions, it is difficult to imagine relaxing spatial regions onto different parts of the $\{ \phi, \chi \}$ potential landscape (at least for potentials with $a$ near $a_{crit}$) without causing $\chi$ to roll off to infinity in the regions between vacua of $\phi$.

%%%%%%%%%%%%%%%%%%%%%%%%%%%%%%
\section{Stability analysis of bubbles}\label{sec:bubbles}

\subsection{Analytical results}

\begin{figure}
\includegraphics[scale=.8]{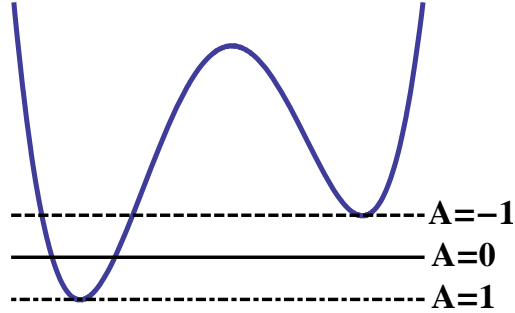}
\caption{ A cross-section of the potential $V_0 + \Delta V$ at $\chi = \chi_{\rm min}$ for various values of $A$. The horizontal lines indicate the zero of the potential for different values of the parameter $A$.
 \label{fig:Aplot}
}
\end{figure}

If we add a symmetry breaking term to the potential in Eq.~\ref{eq:potential}:
\begin{equation}
\Delta V = \delta \mu_{\chi}^4  e^{-3 \chi / M_4} \left( \frac{\phi}{M_{\phi}} + A \right)  
\end{equation}
then the vacua at $\phi = \pm M_{\phi}$ have different energy, whose sign is determined by the constant $A$:
\begin{eqnarray}
A &\geq& 1, \ \ \ V(\pm M_{\phi}) \geq 0, \nonumber \\
-1 &<& A < 1, \ \ \ V(+ M_{\phi}) > 0, V(- M_{\phi}) <0, \nonumber \\
A &\leq& -1, \ \ \ V (\pm M_{\phi}) \leq 0. 
\end{eqnarray}
A cross-section of the potential at $\chi=\chi_{\rm min}$ is shown in Fig.~\ref{fig:Aplot}. The energy in the  decompactified phase sets the zero of the potential.

When the vacua on either side of a domain wall are not degenerate, finite energy solutions correspond to closed membranes separating the low-energy (true vacuum) phase from a background of the high-energy (false vacuum) phase. The lowest energy configurations have spherical symmetry (in addition, deformed bubbles become more spherical as they expand~\cite{Adams:1989su}). We therefore assume spherical symmetry (and again, neglect gravitational effects), leading to the equations of motion
\begin{equation}\label{eq-sphericalevolve}
\partial_t^2 \phi - \partial_r^2 \phi - \frac{2 \partial_r \phi}{r}   = - \frac{dV}{d\phi}, \ \ 
\partial_t^2 \chi - \partial_r^2 \chi - \frac{2 \partial_r \chi}{r} = - \frac{dV}{d\chi} ,
\end{equation}
where $r$ is the radial coordinate perpendicular to the domain wall. 

Zero-energy configurations exist when the bubbles expand from rest with constant acceleration, in which case the spacetime possesses SO(3,1) symmetry. The orbits of the symmetry are characterized by $\rho^2 \equiv r^2 - t^2$. The equations of motion in terms of this coordinate are given by
\begin{equation}\label{eq-SO(3,1)}
\partial_{\rho}^2 \phi + \frac{3 \partial_{\rho} \phi}{\rho}   = \frac{dV}{d\phi}, \ \ 
\partial_{\rho}^2 \chi + \frac{3 \partial_{\rho} \chi}{\rho} = \frac{dV}{d\chi}.
\end{equation}
These coordinates only cover the region of spacetime in the vicinity of the bubble wall. It is necessary to analytically continue across the light cone $r=t$ (where $\rho = 0$) to describe the bubble interior. In this region, the surfaces of constant field trace out the spacelike hyperbolas defined by $\rho^2 \equiv t^2 - r^2$, corresponding to the Milne slicing of Minkowski space.

To obtain a bubble of true vacuum embedded in a sea of the false vacuum, boundary conditions must be chosen such that the fields at $\rho = 0$ are in the basin of attraction of the true vacuum, while the fields at $\rho \rightarrow \infty$ are in the false vacuum. To avoid a singular term in the equations of motion Eq.~\ref{eq-SO(3,1)}, the derivatives of the fields at $\rho = 0$ must be zero: $\{ \partial_{\rho} \phi (0) = 0, \partial_{\rho} \chi (0) = 0 \}$. Just as for the static domain walls, there are two possible classes of solutions: one interpolating between $\chi_{\rm min}$ and the decompactified phase and the other interpolating between the vacua at $\phi \simeq \pm M_{\phi}$ and finite $\chi$.

\begin{figure}
\includegraphics[scale=.8]{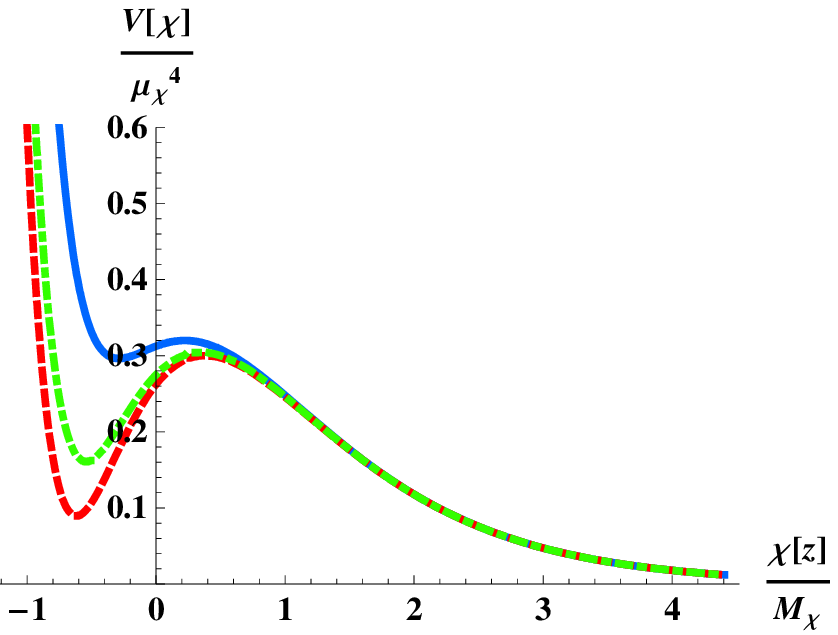}
\includegraphics[scale=.8]{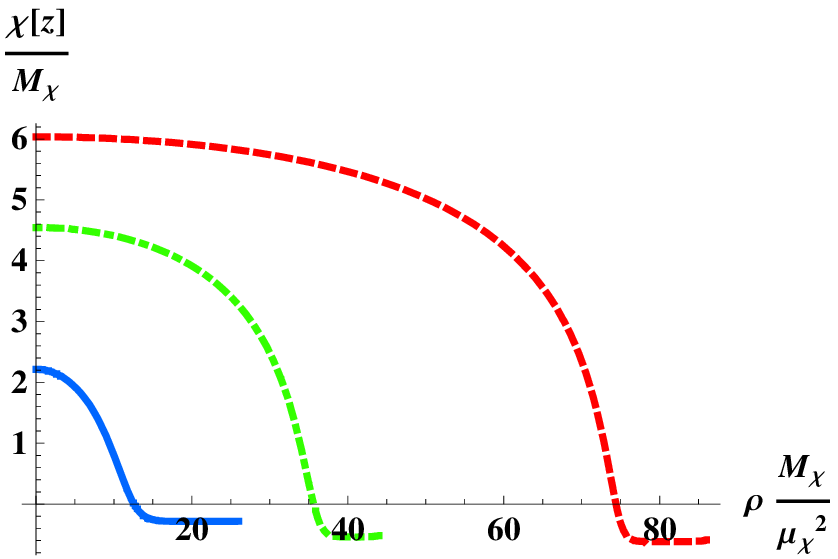}
\caption{Bubbles of the decompactified phase. The bubble wall profile (right panel) for a range of potentials $V_1 (\chi) + \Delta V$ (left panel). As the height of the minimum increases  (red dashed to dot-dashed green to solid blue), the excursion in field space for $\chi$ becomes more and more limited. 
 \label{fig:chibubbles}
}
\end{figure}

We begin by describing bubbles containing the decompactified phase $\chi \rightarrow \infty$ in a background where $\chi = \chi_{\rm min}$ and $\phi \simeq \pm M_{\phi}$. In this case, the bubble wall interpolates between the minimum of the potential $V_1 (\chi)$ and some point to the right of the potential maximum. A number of numerically generated wall profiles and their corresponding potentials are shown in Fig.~\ref{fig:chibubbles}. The range in field space over which $\chi$ can travel is limited by the approximate conservation of energy in the mechanical analog of a particle evolving in the potential $-V_1$. As the relative importance of the symmetry breaking term increases, the local minimum rises. This decreases the available energy, and thus the $\chi$ field has an endpoint closer to the local maximum of the potential.

With the addition of a symmetry breaking term, the vacua at $\phi \simeq \pm M_{\phi}$ are no longer degenerate, and we can search for SO(3,1) invariant bubbles containing $\phi = - M_{\phi}$ embedded in a background where $\phi = + M_{\phi}$. Just as for the domain walls of the previous section, both the $\phi$ and $\chi$ fields have a non-trivial profile inside the wall. In the limit where $\delta \ll 1$, the vacua are nearly degenerate, and we can work in what is known as the thin-wall approximation. In this limit, the  friction term in Eq.~\ref{eq-SO(3,1)} causes the fields to undershoot the false vacuum unless they loiter very near to the true vacuum minimum for an extended range in $\rho$. The initial radius of the bubble (set by the loitering ``time") is therefore very large compared to the thickness of the wall (set by the barrier-crossing ``time"). Additionally, the equations of motion during the traversal from the true to the false vacuum become identical to the domain wall equations studied in the previous section (Eq.~\ref{eq-planarevolve}).

Because the equations of motion reduce to those of the domain wall in the thin-wall limit, many of the results from Sec.~\ref{sec:domainwalls} carry over (with $z$ replaced by $\rho$). Again, we can artificially freeze $\chi$ at $\chi^*$, and then study the backreaction of the bubble wall on the $\chi$ sector. The bubble wall profile for $\phi$ is given approximately by
\begin{equation}\label{eq:thinwallprofile}
\phi_{B} (z) = M_{\phi} \tanh \left[ \frac{\sqrt{2} \mu_{\phi}^2}{M_{\phi}} e^{-n \chi^* / 2 M_{\chi}} ( \rho - \rho_0 ) \right],
\end{equation}
where the initial radius of the bubble is given by
\begin{equation}\label{eq:bubbleradius}
\rho_0 = \frac{3 \sigma}{\Delta V}.
\end{equation}
The tension is identical to that of the domain wall, Eq.~\ref{eq:walltension}, and the difference in vacuum energy at each minimum in our toy model is given approximately by
\begin{equation}
\Delta V = 2 \delta \mu_{\chi}^4 e^{-3 \chi_{\rm min} / M_{\chi}}.
\end{equation}

Accounting for the dynamics of the $\chi$ field, there is one important difference between the bubble and domain wall solutions. Because of the friction term in the equations of motion and non-zero height of the minimum, the excursion of $\chi$ on either side of the bubble wall is limited. This was evident in our construction of the bubbles interpolating between $\chi_{\rm min}$ and the decompactified phase; see Fig.~\ref{fig:chibubbles}. For bubble solutions, it is therefore no longer possible to tune the kick size by placing the wall at arbitrarily large $\chi^*$. This effect is most dramatic outside of the thin wall approximation.

The relevant test for a stable SO(3,1) bubble solution is to compare the rolling velocity in the mechanical analog to the size of the kick, just as for the domain walls. Because of the limited range in $\chi^*$, the parameter $n$ does not determine the stability of the bubble. Instead, as we summarize in Table~\ref{tab:bubblebounds}, comparing the kick size to the barrier height (Eq.~\ref{eq:maximumcondition}) is a good method for determining if a solution exists. The critical value of $a = a_{\rm crit}$ below which stable solutions do not exist should be nearly identical to that of the domain wall solutions. When a solution cannot be found, the $\chi$-field again overshoots its minimum, rolling off to $\chi \rightarrow -\infty$ in finite $\rho$. 

\begin{table*}[tb]
\begin{tabular}{l l l}
\hline
\hline
Uplifted, asymmetric potential,  $a>a_{\rm crit} $ && Stable\\
Uplifted, asymmetric potential,  $a \leq a_{\rm crit} $ && Unstable\\
\hline
\hline
 \end{tabular}
\begin{center}
 \caption{The stability of bubble walls is determined by the value of the parameter $a$. \label{tab:bubblebounds}}
 \end{center}
\end{table*}

Including gravity, the spacetime region in the vicinity of the bubble wall is described by the metric
\begin{equation}\label{eq:gravity31metric}
ds^2 = d\tau^2 + a(\tau)^{2} dH_3^2,
\end{equation}
where $dH_3^2$ is the metric on a hyperboloid with spacelike norm. Inside the bubble, the metric is composed of hyperboloids with timelike norm. The equations of motion for the fields $\{ \phi(\tau), \chi(\tau) \}$ and scale factor $a(\tau)$ are given by
\begin{equation}\label{eq:fieldwgrav}
\ddot{\chi} + 3 \frac{\dot{a}}{a} \dot{\chi} = \frac{dV}{d \chi},  \ \ \ddot{\phi} + 3 \frac{\dot{a}}{a} \dot{\phi} = \frac{dV}{d \phi}, \ \ \dot{a}^2= 1+ \frac{a^2}{3 M_4^2} \left( \frac{\dot{\phi}^2}{2} + \frac{\dot{\chi}^2}{2} - V \right),
\end{equation}
where the dots indicate derivatives with respect to $\tau$. The scale factor evolves over a compact range in $\tau$ between two zeros, each of which correspond to null surfaces. When the derivatives of both $\phi$ and $\chi$ are zero at both ends of the range in $\tau$, the zeros of $a$ are merely coordinate singularities~\cite{Coleman:1980aw,Banks:2002nm}. If not, then the friction terms in the equations of motion cause a curvature singularity to develop~\cite{Cvetic:1994ya,Johnson:2008vn}. This will be the case when the size of the kick overwhelms the rolling velocity, and the $\chi$ field is forced to overshoot its minimum. For such solutions, there is no region of true vacuum inside of the bubble. Instead, the expanding bubble wall contains an expanding spherical timelike singularity. 

The field dynamics we studied above are valid when the radius of the bubble (defined in the thin-wall approximation by Eq.~\ref{eq:bubbleradius}) is much smaller than the Hubble radius defined by the energy of the false vacuum. When gravitational effects are important, there is good reason to expect that stable non-singular solutions are even harder to find. In this limit, the Hubble friction terms in Eq.~\ref{eq:fieldwgrav} become important, causing the field to roll in the steepest direction (this is similar to what happens in slow-roll inflation), which inside of the $\phi$ wall is towards $\chi \rightarrow -\infty$. The same is true for thick walls in the absence of gravity. The result is again a timelike singularity.

\subsection{Vacuum transitions}

Analytically continuing Eq.~\ref{eq:gravity31metric}, the Coleman--de Luccia instanton is obtained. This solution to the Euclidean equations of motion (Eq.~\ref{eq:fieldwgrav}) mediates the nucleation of true vacuum bubbles from a false vacuum phase \cite{Coleman:1980aw}. The singular bubble solutions studied above do not analytically continue into finite action Euclidean instantons. Therefore, the presence of the $\chi$-field can cause the rate for certain transitions out of the false vacuum to go to zero, as first noted in \cite{Johnson:2008vn}. The nucleation of bubbles containing the decompactified phase, which yield finite action instantons, are always allowed.

The lack of certain O(4)-invariant instantons does not preclude the existence of an instanton with less symmetry. For example, O(3)-invariant instantons~\cite{Linde:1980tt,Aguirre:2005nt} can create a localized region of true vacuum. However, if the underlying potential does not allow for a stable bubble wall, then as we show in the next section, our numerical results indicate that when the vacuum energy of the true vacuum is non-negative, the entire bubble interior goes over to the decompactified phase. The only processes that can evolve to a configuration which at late times contains multiple vacua would seem to be a horizon-sized fluctuation out of the false vacuum (larger than both the true and false vacuum horizon). This is similar to the Hawking-Moss transition~\cite{Hawking:1981fz} or the nucleation of a false vacuum bubble~\cite{Lee:1987qc}, which must be larger than the horizon size in order to overcome the encroaching true vacuum exterior. It is unclear how to interpret such events, or even whether they occur at all. In the absence of such fluctuations, this implies that lasting regions of some vacua will not be created by vacuum transitions.

In the physical examples of Sec.~\ref{sec:EMandtypeIIB}, we will consider analogs of the Brown--Teitelboim~\cite{Brown:1988kg} mechanism for membrane nucleation. In these models, there is a charged fundamental brane (the thin-wall limit of the solitonic $\phi$ walls in our toy model) that inherits a coupling to the $\chi$-field. The stability analysis is identical to that presented above, since the branes are characterized by their tension, which in turn determines the effect on the $\chi$ field. Further, in a background field, we expect there to be some process that creates these charged objects, described by a thin-wall CDL or O(3)-invariant instanton. Such a process would be analogous to the Schwinger pair production of electromagnetic charges in an electric field. Again, if the wall is unstable, the resulting configuration will not contain a region of true vacuum at late times, just as in our toy model. 

\subsection{Numerical results}\label{sec:numericsbub}

In order to investigate spherically symmetric solutions that do {\em not} have a boost symmetry, we again employ a set of numerical simulations. The relevant equations of motion are given in Eq.~\ref{eq-sphericalevolve}. We focus on potentials with $\delta \ll 1$, where the thin-wall approximation holds. In this case, as discussed above, the shape of the field profile can be approximated by the domain wall configuration found from the potential with $\delta = 0$. We therefore expect to find similar run-away behavior for $a < a_{\rm crit}$, although the development of the instability will be more involved because of the potentially non-zero energy in the true vacuum.

\begin{figure}
\includegraphics[width=15cm]{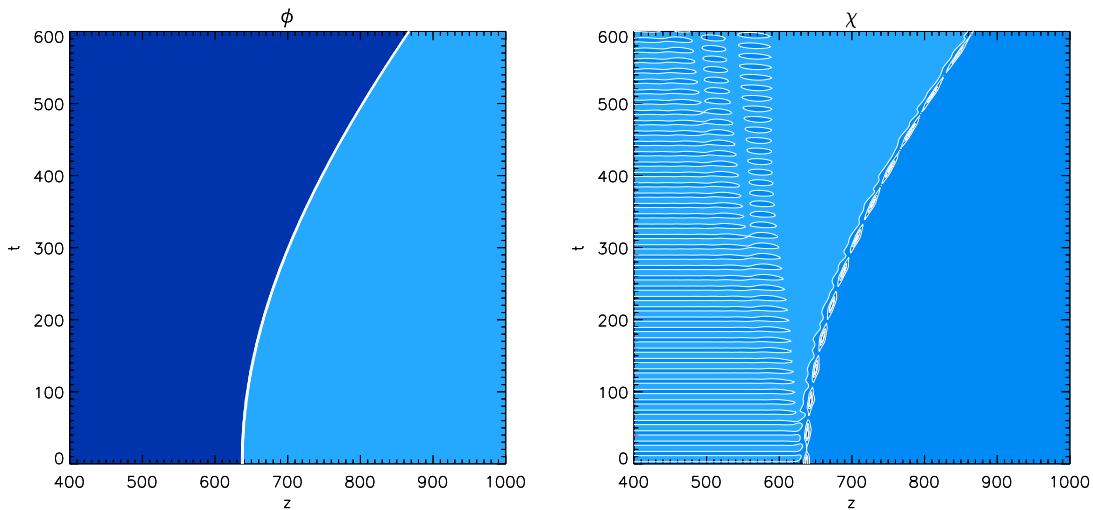}
\caption{The evolution of $\phi$ (left panel) and $\chi$ (right panel) for a stable bubble configuration in the $\{ r, t \}$ plane with a potential specified by the parameters $\{A= 1.0, a = 0.19, \delta = 0.05, n=0.5\}$. The panels show a cross section of the true vacuum bubble, which is separated from the surrounding false vacuum by an expanding domain wall. 
 \label{fig:stablebubble}
}
\end{figure}

Choosing a particular potential, we first set $\delta = 0$ and find the improved domain wall configuration using the methods described in Sec.~\ref{sec:DWnumeric}. We then translate this solution, such that the center 
of the wall is located at a radius given by Eq.~\ref{eq:bubbleradius}. This is used as the initial condition in a simulation with $\delta \ll 1$, and evolves smoothly into an expanding bubble solution. One such stable bubble is shown in Fig.~\ref{fig:stablebubble}. 

Because the symmetry-breaking term changes the vacuum structure of the potential slightly, the improved domain wall configuration we input for $\chi$ does not initially interpolate exactly between the potential minima. This introduces oscillations about the false vacuum, which are apparent in the numerical plots shown in this section. These disturbances are a small perturbation, and will not affect the conclusions of our analysis.

Decreasing $a$ towards the critical value and repeating this procedure, the bubble configuration becomes unstable, seeding a region where the $\chi$ field rolls off to the decompactified phase. It is difficult to find the exact SO(3,1)-invariant profile for stable bubbles, and thus difficult to determine the sharp value of $a_{\rm crit}$, although it is close to that of the corresponding planar domain walls. The sensitivity to small departures from the exact profile indicates that walls are not stable to small perturbations for potentials with $a$ near $a_{\rm crit}$. 

\begin{figure}
\includegraphics[width=15cm]{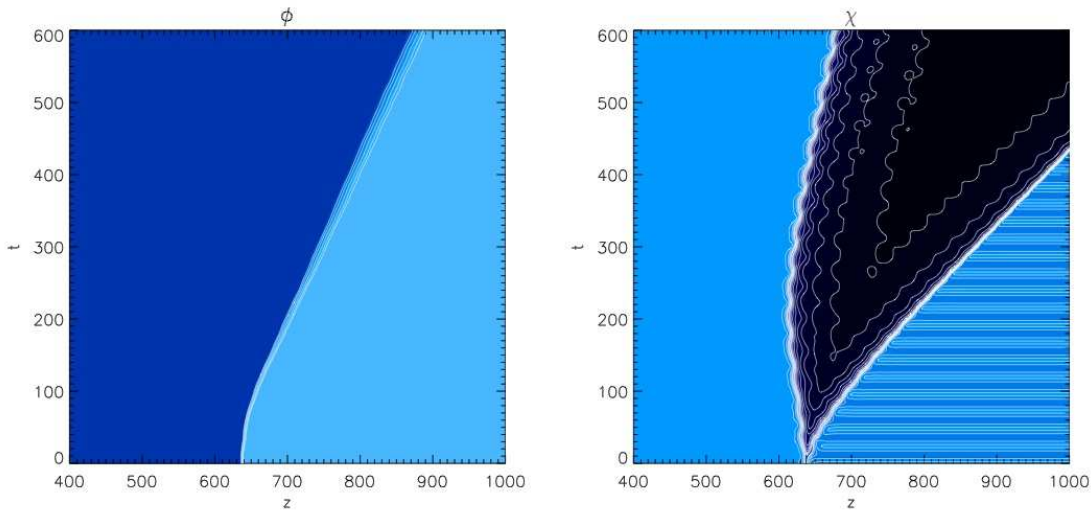}
\caption{The evolution of $\phi$ (left panel) and $\chi$ (right panel) for a bubble configuration in the $\{ r, t \}$ plane with the parameters $\{A= 0.1, a= 0.18, \delta=0.1, n=0.5  \}$. The unstable wall creates a region in which $\chi$ rolls off to $+\infty$ (black), as shown in the right panel. Since this is a higher energy phase than the negative energy bubble interior, the true vacuum region grows, leading to a nested configuration with a true vacuum core (light blue), surrounded by a region where $\chi \rightarrow \infty$ (black), surrounded by the false vacuum (light blue).
 \label{fig:unstableads1}
}
\end{figure}

When it exists, the way that the instability develops depends on the vacuum structure of the potential. If the true vacuum is negative ($A < 1$ in our toy model), it is of lower energy than the decompactified phase at $\chi \rightarrow \infty$. Therefore, if a region of the decompactified phase forms, it will naturally be repelled from the true vacuum interior. An example of this behavior is shown in Fig.~\ref{fig:unstableads1}, where it can be seen that the interface between the true vacuum and the decompactified phase is accelerating out of the bubble interior. Taking a spatial cross section through this configuration, the bubble wall contains an expanding shell of the decompactified phase surrounding a growing true vacuum center.

\begin{figure}
\includegraphics[width=15cm]{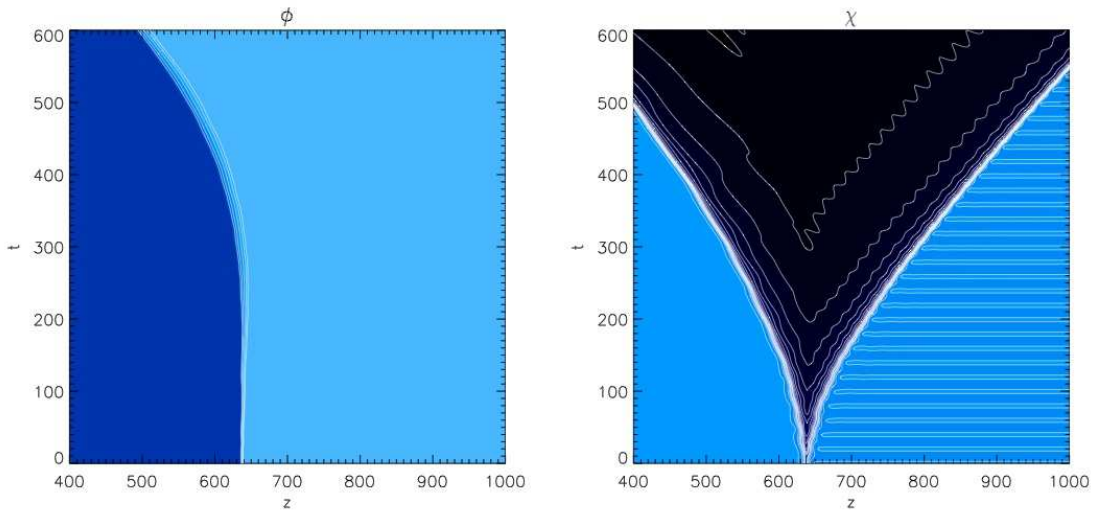}
\caption{The evolution of $\phi$ (left panel) and $\chi$ (right panel) for a bubble configuration in the $\{ r, t \}$ plane with the parameters $\{A= 0.1, a= 0.18, \delta=0.05, n=0.5  \}$. In this case, the reduction in $\delta$ changes the dynamics of the $\phi$-wall in such a way that it implodes into the true vacuum. The region of true vacuum is eventually removed from the spacetime.
 \label{fig:unstableads2}
}
\end{figure}

The dynamics of the instability can also be affected by changes in the behavior of the $\phi$ wall, which lies in the region undergoing decompactification. In the vicinity of the wall, the symmetry-breaking term in the potential is becoming increasingly irrelevant when compared with $V_0$. A decrease in the vacuum energy difference across the $\phi$ wall (as compared to the tension) can change the critical radius for an expanding bubble, and actually reverse the expansion of the wall. An example of this behavior is shown in Fig.~\ref{fig:unstableads2}, which occurs in our toy model for small values of $\delta$. Taking a spatial cross section through the simulation, the bubble wall spawns a shell of the decompactified phase, which then implodes, eventually eating up the entire core of true vacuum. Because the critical radius for expansion diverges as $\chi \rightarrow \infty$, the turn-around of the $\phi$ wall might always occur, although we have not been able to confirm this due to the limitations on the size of our simulation.  

Potentials with a negative energy true vacuum give rise to a big-crunch inside of the bubble~\cite{Abbott:1985kr,Banks:2002nm}. When the wall is unstable, we saw above that the decompactified phase exists in a region separating the bubble interior from the wall. It would be interesting to determine how the singularity develops in the presence of an such an unstable wall.

\begin{figure}
\includegraphics[width=15cm]{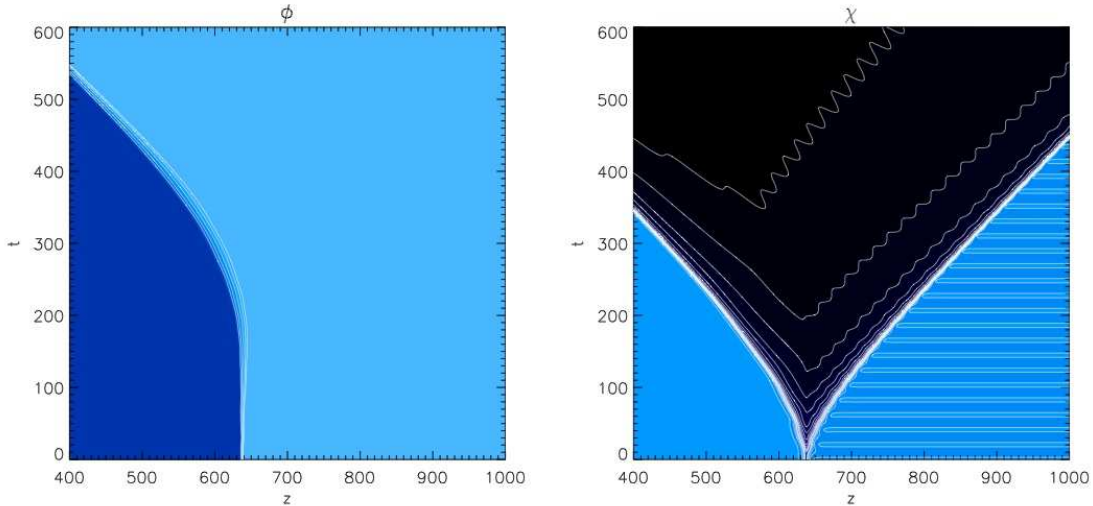}
\caption{Here, we show a contour plot of a configuration for $\phi$ (left) and $\chi$ (right) that initially contains a true vacuum bubble (where $\phi = + M_{\phi}$) embedded in a surrounding false vacuum (where $\phi = -M_{\phi}$). The parameters of the potential are given by $\{A = 1, a = 0.19, \delta = 0.05, n=0.5 \}$. When the phase inside the bubble has zero or positive energy, situations with an unstable wall lead to the entire bubble interior running off to $\chi \rightarrow \infty$ (black), as shown in the right panel. In addition, for this set of parameters, the $\phi$ wall implodes into the bubble.
 \label{fig:unstableds1}
}
\end{figure}

\begin{figure}
\includegraphics[width=15cm]{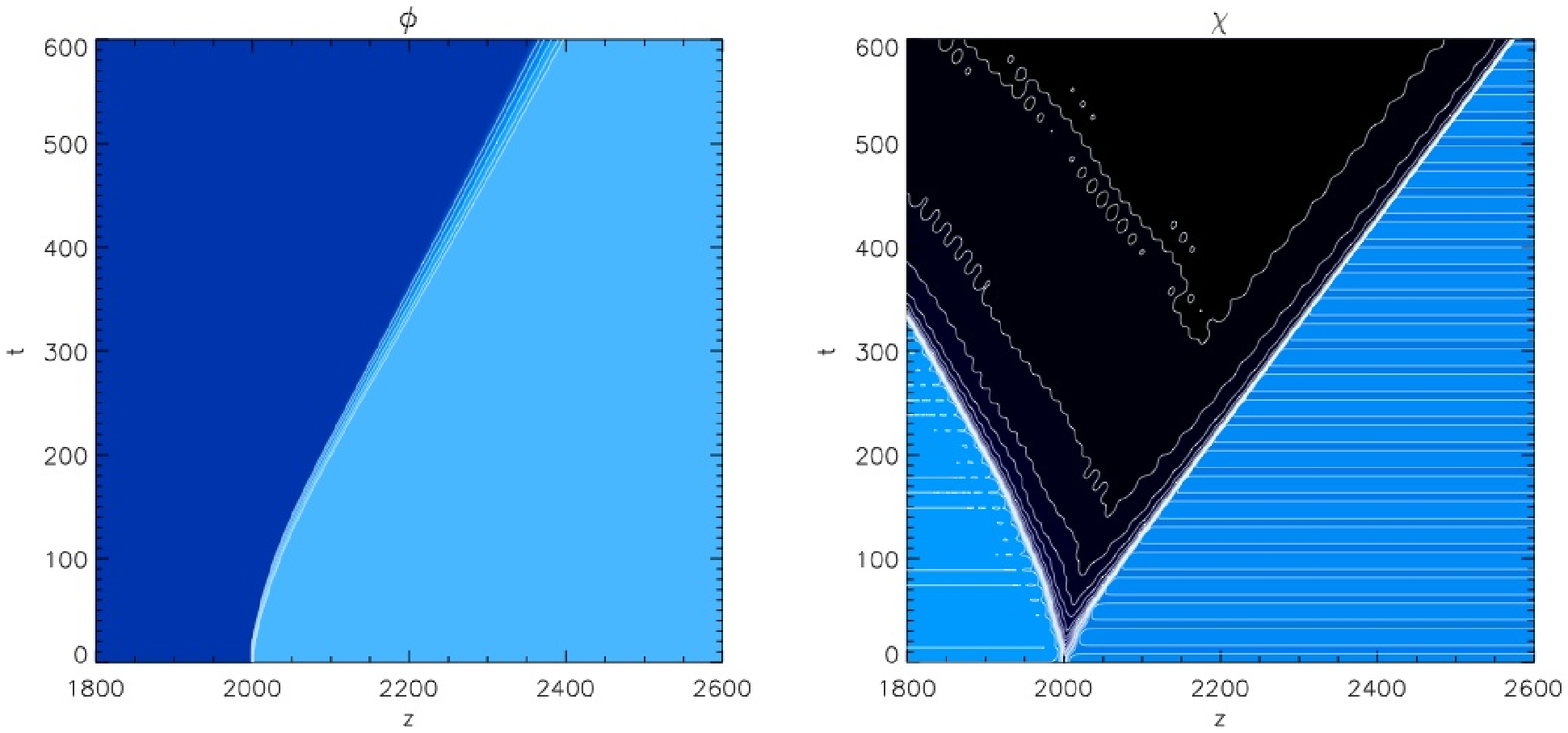}
\caption{A contour plot of a configuration for $\phi$ (left) and $\chi$ (right) that initially contains a true vacuum bubble (where $\phi = + M_{\phi}$) embedded in a surrounding false vacuum (where $\phi = -M_{\phi}$). The parameters of the potential are given by $\{ A = 1, a = 0.22, \delta = 0.05, n=0.5\}$. Using an initial bubble radius much larger than the critical radius, the $\phi$ wall can be made to expand. However, the region where $\chi \rightarrow \infty$ (black region in right panel) still removes the entire true vacuum core from the bubble interior.
 \label{fig:unstableds2}
}
\end{figure}

When the true vacuum has zero or positive energy, it is greater than or equal to the energy of the decompactified phase, and the decompactifying region accelerates into the true vacuum. The dynamics of the $\phi$ wall can also drive the growth of a region containing the decompactified phase. For a bubble near the critical radius, as in Fig.~\ref{fig:unstableds1}, the growing region of the decompactified phase and associated inward moving $\phi$ wall eventually remove the entire region containing the true vacuum. Making the initial radius of the bubble larger, as in Fig.~\ref{fig:unstableds2}, the $\phi$ wall can move outwards (at least for some period of time), but the region containing the decompactified phase again accelerates inwards, removing the region of true vacuum from the spacetime. The results from a number of simulations imply that unstable bubbles separating the vacua at $\phi \simeq \pm M_{\phi}$ always evolve to bubbles containing the decompactified phase.

If we were to include gravity, the Hubble expansion of a positive energy true vacuum interior provides an extra restoring force against the encroaching decompactified phase. If the initial radius of the bubble were comparable to the true vacuum horizon size, and the instability developed on a time-scale comparable to the true vacuum Hubble time, then it is possible (by causality) that the true vacuum phase inside of the bubble is not completely eaten up. If the bubble radius were larger than the true vacuum horizon size, then the bubble interior is guaranteed to survive. However, since the construction of such configurations would require acausal correlations, it is not clear that they are physically well-defined.

%%%%%%%%%%%%%%%%%%%%%%%%%%%%%%%%%%%%%%%%%%%%%%%%%%%%%%%%%%%%

\section{Colliding bubbles with dilatonic couplings}\label{sec:collisions}

In the numerical examples of the previous sections, we have seen that domain walls in compactified theories can be relatively fragile. An additional perturbation to consider, which is very relevant in the context of eternal inflation, is the collision between bubbles. Although the phase transition from the false to the true vacuum does not complete if inflation is eternal, there are an infinite number of collisions with any given bubble. This takes on direct observational relevance because within the picture of eternal inflation, our observable universe could lie within one such bubble, and collisions might leave signatures in cosmological probes such as the CMB (for a review on the observability of bubble collisions, see Ref.~\cite{Aguirre:2009ug}).

If eternal inflation is driven by a potential landscape obtained from a theory with extra dimensions, then the volume modulus can play a dynamical role in the collision between two bubbles. Each bubble is ``dressed" with a non-vacuum configuration of the volume modulus, as we found in the toy model of Sec.~\ref{sec:bubbles}. These lumps of the volume modulus are accelerated with the bubble wall, and since the potential is non-linear, they interact upon collision (this was also suggested in Ref.~\cite{Yang:2009wz}). Bubble collisions can cause a very large perturbation to the configuration of the volume modulus in the collision region, possibly leading to decompactification. This is a generalization of the ``classical transition" mechanism of Ref.~\cite{Easther:2009ft} to multi-field potentials. 

We assess the ability of collisions to seed decompactification by estimating the energy released when two wave-packets of the volume modulus collide. The energy density in each of the wave-packets will be approximately given by the gradient energy estimated in Sec.~\ref{sec:DWanalytic}. When the bubbles are large, they are approximately planar at the location of the collision, and transforming to the frame where the wall is expanding, we can replace $\partial_z \chi \rightarrow \gamma(t) \partial_z \chi$ where $\gamma(t)$ is the Lorentz factor of the expanding wall. Using energy conservation, as in Eq.~\ref{eq:kickcompare}, the gradient energy associated with each of the wave-packets of $\chi$ in the center of mass frame is of order
\begin{equation}
\gamma^2 \left(\partial_z \chi \right)^2 \sim \gamma^2 V_1 (\chi^*).
\end{equation}

When the wave packets attached to each wall collide, some fraction $f$ of the energy density thermalizes due to the anharmonic nature of the potential. Both the importance of the anharmonic terms and the gradient energy increase with $\chi^* - \chi_{\rm min}$. It is also possible that there is energy transfer between the $\phi$ and $\chi$ fields, since they are coupled. If the energy density thermalized is greater than the barrier height, then the collision has the ability to cause decompactification. Neglecting gravity, the condition for no decompactification is
\begin{equation}\label{eq:collisionprediction}
\gamma < \sqrt{\frac{V_1 (\chi_{\rm max})}{f V_{1} (\chi^*)} },
\end{equation}
where $f$ is the fraction of collision energy that thermalizes. The center of mass energy of the collision increases with initial separation, and in the limit where the initial size of the bubble $\rho_0$ is much smaller than the separation $s$, the lorentz factor can be written as $\gamma = s / \rho_0$. The condition Eq.~\ref{eq:collisionprediction} therefore amounts to a restriction on the kinematics of the collision.

Alternatively, using Eq.~\ref{eq:kicksize}, Eq.~\ref{eq:walltension}, noting that $\sigma_{\chi} \sim \sqrt{V_1 (\chi_{\rm max})} M_{\chi}$, and neglecting factors of order one, we can re-write the condition for no decompactification in terms of the tensions of the $\phi$ and $\chi$ walls:
\begin{equation}
\gamma < \frac{\sigma_{\chi} }{ f^{1/2} \sigma_{\phi}}.
\end{equation}
In this picture, the collision can supply the energy necessary to transition from a stable $\phi$ wall to $\chi$ walls. This suggests, from a slightly different perspective, that decompactification will result from sufficiently energetic collisions. 

In order to verify that collisions can indeed cause decompactification, we have run a number of numerical simulations. Following e.g. Ref.~\cite{Aguirre:2009ug}, we transform to a set of hyperbolic coordinates appropriate to the SO(2,1) symmetry of the collision spacetime. By imposing reflective boundary conditions (i.e. imposing symmetry in the fields about each boundary), we model the collision between two identical bubbles by colliding an expanding bubble with the boundary of the simulation. The bubble starts at rest and grows from some fixed initial radius. We vary the kinematics of the collision by changing the spatial size of the simulation. 

\begin{figure}
\includegraphics[width=15cm]{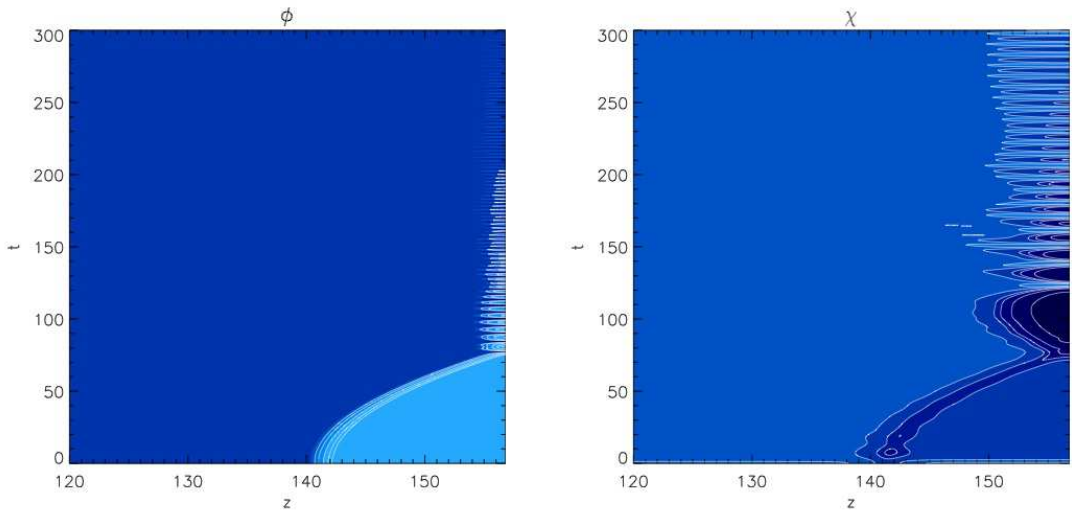}
\caption{The collision between two stable bubbles with $\{a =0.5 , n=0.5, A=2, \delta=0.02 \}$, which does not lead to decompactification in the  future of the collision. The panels show the $\phi$ (left) and $\chi$ (right) evolution of one of the bubbles, which collides with the periodically identified boundary of the simulation. After the collsion at $t \sim 60$, $\chi$ and $\phi$ relax to the true vacuum.
 \label{fig:stablecollision}
}
\end{figure}

Generally, for a wide range in the parameters of the potential, $\{a, n, A, \delta\}$, we find that increasing the separation $s$ between the two colliding bubbles causes the $\chi$ field to get a larger kick to the future of the collision. For some range of kinematics, the collision relaxes back to the true vacuum, and decompactification does not occur. An example is shown in Fig.~\ref{fig:stablecollision}. 

\begin{figure}
\includegraphics[width=15cm]{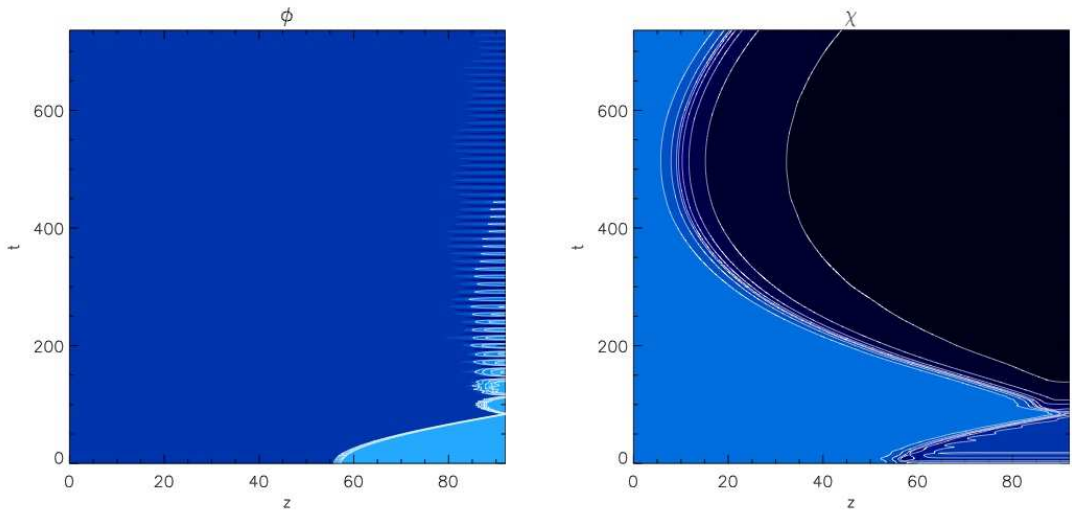}
\caption{The collision between two stable bubbles with $\{a=0.5 , n=0.5, A=1, \delta=0.05 \}$, which results in decompactification in some finite region to the future of the collision. The panels show the $\phi$ (left) and $\chi$ (right) evolution of one of the bubbles, which collides with the periodically identified boundary of the simulation.  After the collsion at $t \sim 100$, $\chi$ gets kicked to the runaway region of $V_1$ in a localized region. This pocket of the decompactified phase (black) expands, but ultimately re-contracts.
 \label{fig:collision_noaccel}
}
\end{figure}

\begin{figure}
\includegraphics[width=15cm]{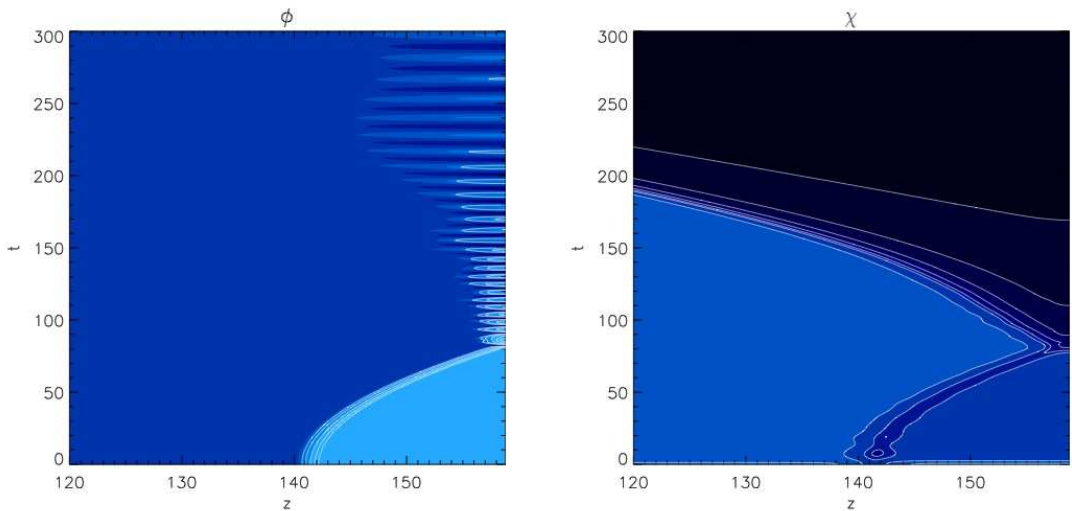}
\caption{The collision between two stable bubbles with $\{a =0.5 , n=0.5, A=2, \delta=0.02 \}$, which results in decompactification in the entire future of the collision. The panels show the $\phi$ (left) and $\chi$ (right) evolution of one of the bubbles, which collides with the periodically identified boundary of the simulation. After the collsion at $t \sim 75$, $\chi$ rolls off to infinity (black).
 \label{fig:collision_accel}
}
\end{figure}

Increasing the separation between the colliding bubbles, we observe two different behaviors depending on the value of $A$. If $A$ is close to one, so that the true vacuum is nearly zero energy, then as shown in Fig.~\ref{fig:collision_noaccel}, a large pocket of the decompactified phase forms and grows, but eventually re-collapses. Increasing $A$, there is a separation past which a growing pocket of the decompactified phase is formed, which then accelerates into the bubble. An example is shown in Fig.~\ref{fig:collision_accel}. The energy of the true vacuum must be sufficiently large to guarantee that the pocket of decompactified phase accelerates into the bubble. This may render vacua with a small cosmological constant more stable towards decompactification from collisions (note that in our universe, the relevant energy scale is that of inflation, and not the present day cosmological constant).

Quantitatively, we can compare our numerical results to the prediction Eq.~\ref{eq:collisionprediction}. Running a number of simulations, we find that Eq.~\ref{eq:collisionprediction} accurately predicts the lorentz factor necessary to produce decompactification for an efficiency somewhere between $0.1 \alt f \alt  0.5$.

In flat space, the bubble separation can increase without bound, and there is always a set of kinematics that leads to decompactification, no matter how inefficient the thermalization process is. In a de Sitter (dS) background, the typical separation for collisions is of order the maximum allowed by causality~\cite{Aguirre:2007wm}, $s \sim H_F^{-1}$. This sets an upper bound on the achievable $\gamma$ factor, of order $\gamma \sim (H_F \rho_0)^{-1}$. Thus, the ability of a collision to decompactify is limited by the size of $H_F \rho_0$, which also parametrizes the importance of gravitational effects in the formation and evolution of bubbles. When gravitational effects can be neglected, the collisions will in many cases be sufficiently energetic to cause decompactification.

Our cosmology cannot exist to the future of collisions that cause decompactification in a region that accelerates into its surroundings. This simple fact has very important implications for assessing the feasibility of observing cosmic bubble collisions during eternal inflation. If decompactification is a fairly generic outcome of collisions, then there will be restrictions on the strength of possible signatures and the number of collisions we might expect to be in our past. Given that the primary motivation for considering landscapes of vacua that drive eternal inflation is the possible existence of extra dimensions, this effect is quite important, and deserves further study in the context of specific models. 

As an interesting aside, there is one sense in which we could exist to the future of a decompactification-inducing bubble collision. If eternal inflation were to occur in three dimensions, then bubble collisions might seed decompactification into a four-dimensional universe. There may be interesting observational consequences of such a scenario.

%%%%%%%%%%%%%%%%%%%%%%%%%%%%%%%%%%%

\section{Dilatonic domain walls in compactified theories}\label{sec:EMandtypeIIB}

In this section, we describe the link between the toy models and more realistic physical models with compatified dimensions.  In particular, we describe how dilatonic couplings between fields arise in theories with extra dimensions. 

After deriving some general properties of these theories, we focus on compactifications of Einstein--Maxwell theory and type IIB string theory. The former compactifications are relatively simple and under good technical control. Compactifications of string theory are more involved, and viable models with four-dimensional dS vacua have only been constructed recently. On the other hand, since string theory is at the moment the best candidate for a theory of quantum gravity, these models are of high interest. 

In both the Type IIB and Einstein--Maxwell examples, we begin with a $D$-dimensional  action of the form 
\begin{equation}\label{eq:Ddimaction}
S =  \frac{M_{D}^{D-2}}{2} \int d^{D} x \sqrt{-\tilde{g}^{(D)}} \tilde{\mathcal{R}}^{(D)} + S_{fields} + S_{sources}.
\end{equation}
Here $M_{D}$ is the $D$-dimensional Planck mass (with $8 \pi G_{D} = M_{D}^{D-2}$), the tilde denotes that each quantity is evaluated in the $D$-dimensional Einstein frame, $S_{fields}$ includes the matter fields present in the theory (e.g. gauge fields and bulk scalars), and $S_{sources}$ includes any branes or other sources that may be present. The fields and sources depend on the theory under consideration, and we defer discussion of these pieces of the action to the particular examples in following sections.

We generally assume that the $D$-dimensional space-time factorizes as $\cM_{D} = \cM_4 \times_w \Omega_q$, where $\Omega_q$ is a compact $q$-dimensional manifold and $w$ indicates that warp factors might be present. The metric is block-diagonal, and in the absence of warping, we will begin with the ansatz: 
\begin{equation}\label{eq:Ddimeinsteinmetric}
ds^2 = \exp \left[ - \sqrt{\frac{2 q}{q+2}} \frac{\chi ({\bf x})}{M_4} \right]  g^{(4)}_{\mu \nu} ( {\bf x} ) dx^{\mu} dx^{\nu} + M_{q+4}^{-2} \exp \left[ \sqrt{\frac{8}{q (q+2)}} \frac{\chi ({\bf x})}{M_4} \right]  g^{(q)}_{m n} dy^m dy^n.
\end{equation}
The field $\chi(x)$ is the volume modulus of $\Omega_q$. There may also be non-trivial warping in different regions of the compact manifold, as will arise in the Type IIB examples below. In this case there are a number of subtleties which we comment on in Sec.~\ref{sec:typeIIB}.  

Inserting this metric ansatz into Eq.~\ref{eq:Ddimaction}, and integrating over the compact space, we obtain 
\begin{equation}\label{eq:einsteinaction}
S = \int d^{4} x \sqrt{ - g} \left[ \frac{M_{4}^2}{2}
\mathcal{R} - 
\frac{1}{2} g^{\mu \nu} (\partial_{\mu}
\chi) (\partial_{\nu} \chi) - V (\chi) \right]  + S_{sources} + \ldots ,
\end{equation}
where
\begin{equation}
M_4 \equiv M_{D}  \left( \int d^q y \sqrt{g^{(q)}} \right)^{1/2} = M_{D} {\rm Vol (\Omega_q)}^{1/2},
\end{equation}
defines the four-dimensional Planck mass. The potential for $\chi$ depends both on the field content of the higher dimensional theory and the assumptions made about the form of the compact manifold. We assume that the dimensional reduction is self-consistent -- namely, that the Kaluza--Klein modes can be integrated out and the volume of the compact manifold is somewhat larger than $M_D^{-q}$.

We now consider a number of examples where the dimensional reduction can be carried out explicitly.

%%%%%%%%%%%%%%%%%%%%%%%%%%%%%%%%%%
\subsection{Einstein--Maxwell theory}

Perhaps the simplest example of a theory possessing dS vacua is the Einstein--Maxwell system with a positive cosmological constant compactified on a $q$-dimensional sphere (recent papers describing this model include~\cite{Carroll:2009dn,BlancoPillado:2009di,Yang:2009wz}). The $D$ dimensional action for the fields in this example is given by
\begin{equation}\label{eq:EMaction}
S_{fields} =  \frac{M_{D}^{D-2}}{2} \int d^{D} x \sqrt{-\tilde{g}^{(D)}} \left[ - 2 \Lambda - \frac{1}{2 q!} \tilde{F}_q^2 \right],
\end{equation}
where $F_q$ is a $q$-form magnetic flux wrapping the compact manifold (equivalently, we could add the dual $4$-form electric field strength), and $\Lambda > 0$. The field equations for the $q$-form are satisfied for
\begin{equation}
\tilde{F}_q = Q \sin^{q-1} \theta_1 \ldots \sin \theta_{q-1} d\theta_1 \ldots \wedge d\theta_q.
\end{equation}

\begin{figure}
\includegraphics[scale=.8]{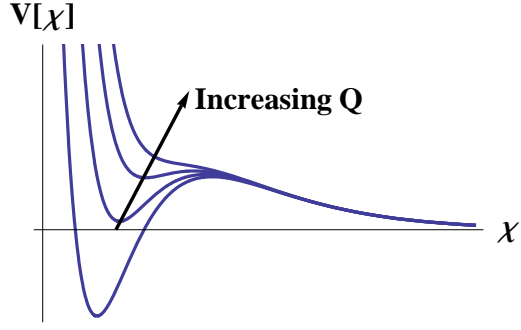}
\caption{A sketch of the potential Eq.~\ref{eq:radionpotential} for fixed $\Lambda$ and various $Q$.
 \label{fig:EMpotential}
}
\end{figure}

In the dimensionally reduced theory, these sources give rise to a potential for $\chi$ given by
\begin{eqnarray}\label{eq:radionpotential}
V (\chi) &=& 
\frac{M_{4}^4}{2 {\rm Vol(S^q)}} 
\left[ - q (q-1) e^{-
2/\lambda(q) \frac{\chi}{M_{4}}} 
+ \frac{2\Lambda}{M_{q+4}^2} e^{- \lambda(q)
\frac{\chi}{M_{4}}}
 + \frac{Q^2}{2} e^{ - 3 \lambda(q) \frac{\chi}{M_{4}}}  \right],
\end{eqnarray}
where the first term is due to the positive curvature of the compactification manifold, we have introduced $\lambda(q)=\sqrt{\frac{2 q}{(q+2)}} $ to simplify the expressions, and ${\rm Vol} (S^{q})$ is the volume of a unit $q$-sphere
\begin{equation}
{\rm Vol} (S^{q}) = \frac{2 \pi^{(q+1) / 2}}{\Gamma \left( \frac{q+1}{2} \right)}.
\end{equation}
Depending on the charge, there can exist positive, negative, or zero energy minima of this potential, giving rise to four-dimensional dS, AdS, or Minkowski vacua. A sketch of the potential for various $Q$ is shown in Fig.~\ref{fig:EMpotential}. 

We now turn to $S_{sources}$. A fundamental $m+2$-brane in the $D$-dimensional theory can give rise to a $2$-brane in four dimensions, with $m$ legs of the brane wrapped on a cycle of the compact manifold. For the spherical compactifications under consideration, stable configurations only exist for an unwrapped ($m=0$) or fully wrapped ($m=q$) brane. The $m+2$-brane naturally sources electric or magnetic flux, which can decompose upon dimensional reduction into legs living on the compact portion of the manifold, the non-compact portion of the manifold, or both. For example, a 2-brane can source a $q$-form magnetic flux living on the sphere or a $4$-form electric flux living in the four non-compact dimensions. A wrapped $q+2$-brane sources both a $q$-form flux on the sphere and a $4$-form in the non-compact dimensions. 

Moving across a 2-brane, the effect of the charge is to shift $Q \rightarrow Q - q$. Moving across a wrapped $q+2$-brane, the effect is to shift $\Lambda \rightarrow \Lambda - q$.\footnote{A $D=q+4$-flux, as the one sourced by the $q+2$-brane, acts effectively as a cosmological constant in a $D$-dimensional theory. } This changes the vacuum structure of the potential for $\chi$, and therefore the membranes separate four-dimensional vacua with different vacuum energy (hence, the membranes must be spherical to have finite energy). In analogy with Schwinger production of electron-positron pairs in a constant electric field, spherical charged membranes are nucleated from the background fields~\cite{Brown:1988kg} used to stabilize the volume modulus (see Ref.~\cite{BlancoPillado:2009di} for a recent discussion of membrane production in the Einstein--Maxwell model).

The source action in the $D$-dimensional Einstein frame, including a coupling to the $3+m$-form gauge field ${\bf A}$, is given by 
\begin{equation}\label{eq:actionsourcesbrane}
S_{sources} = - \sigma \int_{\Sigma} \sqrt{\tilde{\gamma}} d^{3+m} \xi + q \int_{\Sigma} {\bf A}, 
\end{equation}
where $\sigma$ is the tension of the brane, $q$ is the charge, and the integrals are over the world-volume $\Sigma$ of the brane; $\tilde{\gamma}$ is the worldsheet metric. 

The dimensional reduction to the four-dimensional Einstein frame is performed using the metric Eq.~\ref{eq:Ddimeinsteinmetric}. In general, the addition of sources deforms the spherical compactification manifold, introducing warping into the metric (see e.g.~\cite{Kinoshita:2009hh} for a discussion of warped compactifications of this type). We neglect this effect, which should be a good approximation in the limit where $Q \gg q$. This also implies that the vacua on either side of the brane are nearly degenerate, and so we focus on the role of the tension, which in the four-dimensional Einstein frame contributes an action
\begin{equation}\label{eq:reduceddomainaction}
 \sigma \int_{\Sigma} \sqrt{\tilde{\gamma}} d^3 \xi d^m \xi  =  M_4^3 \  \left( \frac{\sigma}{M_{D}^{m+3}} \right) \left( \frac{\Pi_m}{{\rm Vol(\Omega_q)}^{3/2} } \right) \int_{\Sigma} \sqrt{\gamma} \exp \left[ - \frac{3q-2 m}{\sqrt{2 q(q+2)}} \frac{\chi ({\bf \xi})}{M_4} \right] d^3 \xi.
\end{equation}
Here $\gamma$ is the worldsheet metric defined with respect to the four-dimensional Einstein frame metric $g^{(4)}_{\mu \nu}$, $\Pi_m$ is the period of the $m$-cycle that the brane wraps (for the purpose of illustration, we have allowed for any number of legs of the brane to be wrapped; for $m=q$, $\Pi_m = {\rm Vol(\Omega_q)}$). The exponent in the coupling to $\chi$ is negative definite since $m \leq q$.

We can now make connection with the toy model described in Section~\ref{sec:toymodel} (which the reader should now recognize as arising from six-dimensional Einstein--Maxwell theory). Due to the tension term in the action, the potential for the volume modulus $\chi$ gets an extra contribution at the location of the brane, inducing a spatial derivative approximated by 
\begin{equation}\label{eq:deltachiEM}
\Delta \chi' = - M_4^2 \left( \frac{\sigma}{M_{D}^{3+m}} \right)  \left( \frac{\Pi_m}{{\rm Vol (S^q)}^{3/2} } \right) \frac{3q-2 m}{\sqrt{2 q(q+2)}}  \exp \left[ - \frac{3q-2 m}{\sqrt{2 q(q+2)}} \frac{ \chi^* }{M_4} \right],
\end{equation}
where the prime denotes a derivative with respect to $z$ or $r$, depending on the assumed symmetry of the wall. As described in Sections~\ref{sec:domainwalls} and~\ref{sec:bubbles}, the existence of static planar or SO(3,1)-invariant solutions including the brane and $\chi$ field depends both on the asymptotics and parameters of the potential for $\chi$. When $\Lambda = 0$, the local minimum is AdS and the potential for $\chi$ approaches zero from below, resembling the AdS type potentials in our toy model. In this case, as we found in Sec.~\ref{sec:domainwalls}, there is always a stable domain wall solution. For $\Lambda > 0$, the local minimum can be dS and the potential approaches zero from above, resembling the Uplifted class of models. For such potentials, there is only a non-singular static or SO(3,1) invariant solution when the spatial derivative induced by the brane is smaller than the possible gradients induced by the potential.  

Applying the first analytic estimate discussed in Sec.~\ref{sec:domainwalls}, we compare the change in the spatial gradient of the field $\chi$ (Eq.~\ref{eq:deltachiEM}) to the height of the potential maximum for $\chi$. The location and height of the maximum are given by~\cite{Carroll:2009dn}
\begin{eqnarray}
\chi_{\rm max} &\simeq& \frac{\sqrt{q(q+2)}}{2\sqrt{2}} \log \left[ (q-1) (q+2) \frac{M_D^2}{2 \Lambda}\right], \\
V(\chi_{\rm max}) &\simeq& \frac{M_4^4}{{\rm Vol(S^q)} (2+q)^{1+q/2} (q-1)^{q/2}} \left( \frac{2 \Lambda}{M_{D}^2} \right)^{1+q/2}.
\end{eqnarray}
Substituting with $\chi^* = \chi_{\rm max}$, the ratio of the kick to the rolling velocity in the potential $V_1$ scales like
\begin{equation}
\frac{(\Delta \chi')}{2 \sqrt{2 V_{\rm max}}} \sim \frac{\Pi_m}{ {\rm Vol} (S^q)}  \frac{\sigma}{M_{D}^{3+m}}  \left( \frac{2 \Lambda}{M_{D}^2}\right)^{(q-m-1)/2}.
\end{equation}
The first factors in this expression are typically order one numbers, since $\Pi_{0,q} = ( 1, {\rm Vol} (S^q) )$ (unwrapped and wrapped cases respectively) and the tension of the branes should scale with $M_{D}$.\footnote{If these are gravity solutions, then the horizon associated with the branes scales with the tension -- sub-planckian horizons would not be under good theoretical control, and we therefore expect that $\sigma \agt M_D^{3+m}$.} In the last factor, we expect $\Lambda/M_{D}^2$ to be small for the semi-classical approximation to be valid. Thus, the magnitude of the ratio depends on the values of $q$ and $m$, i.e. on how the branes wrap the extra dimensions. 

When $q-m-1 > 0$, the kinetic energy is always small compared to the potential energy, and domain walls are expected to be stable. For $q-m-1 < 0$, the ratio is large and it seems difficult to obtain stable domain walls. Maximally wrapped branes are therefore always unstable. Indeed, the asymptotic falloff of the gradient due to the brane is always larger than the asymptotic gradient due to the potential for $\chi$, ensuring that stable solutions are truly impossible to construct. Exceptions to this rule are only possible if the tension is tuned to be sufficiently small compared with $M_{D}$, which seems difficult to obtain for gravitating branes that can be described by the Einstein--Maxwell model. 

From this example, we infer that when a brane wraps some internal cycle, it is more difficult to find stable domain walls. This is due to the fact that the brane tension in the four-dimensional Einstein frame falls off more slowly with increasing $\chi$ if the brane has one or more legs in the compactification manifold. In a model where the domain walls are unstable, their nucleation cannot proceed via the standard O(4)-invariant instanton. However, one might expect there to be some process by which membrane production can still occur   via some O(3)-invariant~\cite{Linde:1980tt,Aguirre:2005nt} or constrained instanton~\cite{Affleck:1980mp}. In this case, once formed, if the interior of the membrane has zero or positive energy density it decompactifies as described in Sec.~\ref{sec:numericsbub}. Therefore, some vacua cannot be populated by membrane nucleation in the Einstein--Maxwell model.

%%%%%%%%%%%%%%%%%%%%%
\subsection{Compactifications of type II string theory}\label{sec:typeIIB}

There exist a multitude of compactifications of string theory to four dimensions. M theory, heterotic and type II string theories can all result in effectively four-dimensional models.\footnote{A short survey of the phenomenological benefits and shortcomings of various models can be found in \cite{Denef:2008wq}.} However, matching these models with phenomenological requirements is non-trivial. In particular, obtaining four-dimensional dS vacua has proven to be difficult. In this section, we focus on compactifications of type IIB string theory, where currently the most reliable constructions of dS vacua have been performed.\footnote{This analysis can also be relevant for the IIA vacua of \cite{Palti:2008mg}, which, given a restriction on the allowed fluxes, are related to IIB vacua by mirror symmetry.  Other IIA compactifications leading to dS models also exist \cite{Silverstein:2007ac, Haque:2008jz}, see however \cite{Danielsson:2009ff}. More work on the difficulty of constructing type IIA dS vacua, including several no-go theorems, can be found in e.g. \cite{Hertzberg:2007wc, Caviezel:2008tf,Flauger:2008ad}.}  We only recall the main features of these models (readers interested in the details and a more extensive list of references are referred to one of the many reviews on the subject, e.g. \cite{Douglas:2006es, Grana:2005jc, Denef:2008wq}).

Type II string theories contain a wide variety of stable branes, sourcing different higher-dimensional fluxes. Allowing these fluxes and branes to pierce and wrap cycles in a compact manifold can stabilize the shape and size of this manifold. For type IIB compactifications we can choose the internal manifold to be conformally Calabi--Yau, in which case it is described by complex structure and K\"ahler moduli. The volume modulus $\chi$ defined in Eq.~\ref{eq:Ddimeinsteinmetric} is a K\"ahler modulus. The complex structure moduli are analogous to the $\phi$ field in the toy model of Sec.~\ref{sec:toymodel}.

As was mentioned in the previous section, the addition of branes and fluxes generally leads to non-trivial warping of the compactification manifold. Warping is of phenomenological interest for generating hierarchies in particle physics~\cite{Randall:1999ee,Randall:1999vf,Giddings:2001yu,DeWolfe:2002nn}, and is a key ingredient required to stabilize moduli in some scenarios~\cite{Kachru:2003aw}. There are a number of subtleties that come with identifying moduli and determining their dynamics in a theory with warping~\cite{Giddings:2005ff,Shiu:2008ry,Douglas:2008jx,Frey:2008xw}. We are mainly interested in the properties of complex structure moduli and the overall volume modulus, and so it is sufficient to use the metric ansatz of Ref.~\cite{Frey:2008xw}:
\begin{equation}\label{eq:warpedmetric}
ds^2 = \frac{e^{2 \Omega(c)}}{\left[ e^{-4 A_0(y)} + c(x) \right]^{1/2}} \left( g^{(4)}_{\mu \nu} (\mathbf{x}) dx^{\mu} dx^{\nu} + 2 \partial_j B \partial_{\mu} c dx^{\mu} dy^{j} \right) +\left[ e^{-4 A_0(y)} + c(x) \right]^{1/2} g^{(6)}_{ij} (\mathbf{y}) dy^i dy^j ,
\end{equation}
where $c$ is the overall volume, related to $\chi$ by
\begin{equation}\label{eq:ctochi}
c (x) = \exp \left[ \sqrt{\frac{2}{3}} \frac{\chi (x)}{M_4} \right],
\end{equation}
and $A_0(y)$ is a warp factor associated with the metric $g^{(6)}_{ij}$. $B$ is a compensator field introduced to satisfy the Einstein equations, whose form does not play a role in the following, and
\begin{equation}
e^{2 \Omega(c)} = \frac{ \int d^6 y \sqrt{g^{(6)}} }{ \int d^6 y \sqrt{g^{(6)}} e^{-4 A_0} + c(x) \  \int d^6 y \sqrt{g^{(6)}}}.
\end{equation}
The relation between the 4 and 10-dimensional Planck mass is as before:
\begin{equation}
M_4^2 = M_{10}^8 \int d^6 y \sqrt{g^{(6)}} = M_{10}^2  {\rm Vol (\Omega_6)},
\end{equation}
where ${\rm Vol (\Omega_6)}$ is the volume of the compact manifold measured in units of $M_{10}$.

We are interested in two limiting forms of the warped metric Eq.~\ref{eq:warpedmetric}. When there is no warping, or when the warping is everywhere very small compared to $c$, we recover the metric Eq.~\ref{eq:Ddimeinsteinmetric}. Alternatively, if there is strong warping $e^{-4 A_0(y)} \gg c$ in a localized region of the Calabi--Yau, then upon the shift
\begin{equation}
c \rightarrow c -  \frac{ \int d^6 y \sqrt{g^{(6)}} e^{-4 A_0} }{ \int d^6 y \sqrt{g^{(6)}}  },
\end{equation}
and substituting with Eq.~\ref{eq:ctochi}, the metric becomes
\begin{equation}\label{eq:metricstrongwarping}
ds^2 = e^{2 A_0(y)} \exp \left[ - \sqrt{\frac{2}{3}} \frac{\chi (x)}{M_4} \right] \left( g^{(4)}_{\mu \nu} (\mathbf{x}) dx^{\mu} dx^{\nu} + 2 \partial_j B \partial_{\mu} c dx^{\mu} dy^{j} \right) + e^{-2 A_0(y)}  g^{(6)}_{ij} (\mathbf{y}) dy^i dy^j.
\end{equation}

\subsubsection{The scalar potential}
The four-dimensional effective field theory derived from type IIB compactifications on conformal Calabi--Yau manifolds has  $\cN=1$ supersymmetry. It is characterised by a K\"ahler potential $K$ and a flux-induced Gukov--Vafa--Witten (GVW) superpotential $W$ \cite{Gukov:1999ya}, which determine the scalar potential $V$ for the moduli \cite{Giddings:2001yu, DeWolfe:2002nn}. The effective theory has a no-scale structure, implying that the flux-induced potential is independent of all K\"ahler moduli except the volume modulus, which sets the overall scale of the potential. The GVW superpotential and no-scale structure survives the corrections due to warping~\cite{Giddings:2001yu, DeWolfe:2002nn,Frey:2008xw}, which can be encapsulated entirely in the shift of the field $c$ as above. The dependence on the volume modulus $\chi$ is therefore a simple scaling
\be
V = \exp(- \sqrt{6} \chi/M_4) V_0( \bf{\phi}),
\ee
where $\bf{\phi}$ collectively denotes all the flux-fixed fields of the compactification. Thus $\chi$ is a runaway direction in the tree level potential, exactly as in the toy model discussed in Section~\ref{sec:toymodel}. We will see below that the volume modulus also couples to the branes in the theory, just as in the Einstein--Maxwell example, and will obstruct their stability if it is left unfixed. To fix this modulus, sub-leading quantum corrections must be added to the potential. 

Existing mechanisms for moduli stabilization, such as KKLT~\cite{Kachru:2003aw} and BBCQ~\cite{Balasubramanian:2005zx}, take advantage of non-perturbative $g_s$ or perturbative $\alpha'$ corrections that induce a potential for the K\"ahler moduli. In both cases, an AdS minimum can be produced. In the KKLT model, only the $g_s$ corrections are considered. The AdS minimum is supersymmetric and of depth
\be \label{eq:KKLTminheight}
V_{KKLT}(c_{\rm min}) \approx - \frac{a^2 A^2 e^{-2a c_{\rm min}}}{6 c_{\rm min}},
\ee
where $a$ and $A$ are undetermined (but in principle calculable) constants appearing in the non-perturbative corrections to the superpotential $W = W_0 + A e^{a c}$~\cite{Kachru:2003aw}.\footnote{The (possibly large) shift in $c$ made above can be absorbed into the coefficient $A$ as in~\cite{Frey:2008xw}.} Since $c_{\rm min} \sim - \frac{1}{a} \log \left[ \frac{|W_0|}{A}\right]$, and we require $c_{\rm min} \gg 1$, the construction only works for very small, but non-zero, vacuum expectation values of the tree level superpotential $W_0$. In the absence of other sources, the full potential approaches zero from below at large $c$.

In the BBCQ model (a.k.a. LARGE volume compactifications) both $g_s$ and $\alpha'$ corrections are considered. To obtain a vacuum in this model, it is necessary to consider two different K\"ahler moduli, where the overall volume of the compactification is some linear combination of the two fields. For small values of the flux superpotential $|W_0|$, a KKLT minimum is reproduced, whereas for larger $|W_0|$ a non-supersymmetric AdS minimum is obtained at exponentially large volume of depth~\cite{Balasubramanian:2005zx}
\be
V_{BBCQ}(\chi_{\rm min}) \approx - e^{K_{cs}} |A_s W_0| \frac{\chi}{M_4}e^{-3 \sqrt{\frac{3}{2}}\chi/M_4}. 
\ee
The K\"ahler potential $K_{cs}$ and flux superpotential $W_0$ are in principle calculable from the fluxes and vacuum expectation values of the complex structure moduli and axio-dilaton. The constant $A_s$ arises from the non-perturbative corrections to the potential for the K\"ahler moduli. Again, the potential approaches zero from below at large volume. 

As already noted, both KKLT and BBCQ typically lead to AdS vacua, which must then be ``uplifted" to Minkowski or dS vacua by adding extra sources to the theory. This can be achieved by space-filling anti-$D3$ branes localized in a warped throat of the Calabi--Yau \cite{Kachru:2003aw}. Alternatively, one can use space-filling, partially wrapped $D7$-branes with world-volume gauge fluxes~\cite{Burgess:2003ic}. The scaling of the contribution of the anti-$D3$ branes with volume and warping is obtained from the action Eq.~\ref{eq:actionsourcesbrane} using the metric ansatz Eq.~\ref{eq:warpedmetric} in the appropriate limit:
\begin{eqnarray}\label{eq:D3action}
&{\rm strong \ warping:}& \ S_3 =  \int d^4 x \sqrt{g} \ \frac{\sigma_3 e^{4 A_0}}{c^2}, \\ 
&{\rm weak \ warping:}& \ S_3 =  \int d^4 x \sqrt{g} \ \frac{\sigma_3}{c^3},
\end{eqnarray}
where $\sigma_3$ is the brane tension, and we have neglected the effect of the $5$-form flux sourced by the brane (roughly, this contributes a factor of 2~\cite{Kachru:2003aw}).

In the case of a partially wrapped $D7$-brane, the uplifting term in the four-dimensional action is proportional to the integrated world-volume flux~\cite{Burgess:2003ic,Haack:2006cy}, and scales similarly to the contribution from the anti-$D3$ branes. Allowing for the possibility of stacks of branes, these new ingredients add an uplifting term to the potential for the volume modulus of the form:
\begin{eqnarray} \label{eq:vuplift}
&{\rm \overline{D3} \ or \ D7 \ without \ warping:}& \ V_{uplift} = D e^{- \sqrt{6} \chi/M_4}, \\
&{\rm \overline{D3} \ or \ D7 \ with \ warping :}& \ V_{uplift} = D e^{-\frac{2 \sqrt{2}  \chi}{\sqrt{3}M_4}} \nonumber,
\end{eqnarray}
with the constant $D$ dependent on the warping, volume of wrapped cycles, gauge dynamics, and number of branes. These sources change the asymptotics of the potential such that it approaches zero from above, and resemble the Uplifted type of potentials in the toy model of Sec.~\ref{sec:toymodel}. 

In order to obtain a vacuum after uplifting, Eq.~\ref{eq:vuplift} must be comparable in size to the other contributions to the potential at the location of the original AdS minimum (that is, Eq.~\ref{eq:vuplift} must roughly be of order the depth of the AdS minimum). This generally requires the coefficient $D$ to be very small since the uplifting terms fall off much more gradually with increasing $\chi$ (or equivalently $c$) than the other terms in the potential. In the KKLT example, $D$ is made small by locating a stack of anti-$D3$ branes at the bottom of a strongly warped throat. The anti-$D3$ brane action Eq.~\ref{eq:D3action} depends on the warping, which can lead to an exponentially small $D$. For BBCQ, it is possible to use $D7$ branes to uplift the AdS vacua~\cite{Villadoro:2005yq,Haack:2006cy}. In this case there are a number of ways to obtain a small coefficient $D$, including the presence of warping,  features of the internal space, and contributions from the world volume gauge theory~\cite{Burgess:2003ic,Haack:2006cy,Kallosh:2004dv}. It may also be possible for warped throats to exist in the BBCQ setup, opening up the possibility for anti-$D3$ uplifting, although a detailed analysis along the lines of Ref.~\cite{Frey:2008xw} is necessary to determine the viability of such a scenario. 

\subsubsection{Domain Walls}

In string and M theory, there are (possibly wrapped) branes which are charged under the fluxes used to stabilize the moduli of the compactification. Moving across such branes, the fluxes are adjusted, leading to changes in the potential for various moduli. Depending on the compactification setup, this can lead to quite dramatic differences, such as changing the relative importance of various non-perturbative contributions to the potential for the K\"ahler moduli. For simplicity, and since it is a generic effect, we will here assume that the most relevant difference is a (small) change in vacuum energy. 

In Type IIB string theory, there are $D5$ and $NS5$-branes which can wrap internal 3-cycles of a Calabi--Yau compactification manifold and give rise to charged membranes in the four-dimensional effective theory. Again, there are processes by which these charged branes can be nucleated from a background configuration of the fluxes~\cite{Bousso:2000xa,Feng:2000if,Garriga:2003gv}.\footnote{Some caution is needed here, since it has been argued that brane nucleation is an intrinsically stringy process, lying outside of the supergravity approximation~\cite{deAlwis:2006cb,deAlwis:2006am}.}
In a warped throat with $D3$ and anti-$D3$ branes, the nucleation of  wrapped $NS5$ branes can be described by the Kachru--Pearson--Verlinde (KPV) mechanism~\cite{Kachru:2002gs} (see also~\cite{Frey:2003dm,Freivogel:2008wm,Brown:2009yb}). In a more general setting, solitonic domain walls can be constructed between vacua for the flux-stabilized moduli fields~\cite{Johnson:2008vn,Danielsson:2006xw}. It was shown in~\cite{Johnson:2008kc} (see also e.g.~\cite{Dine:2007er} and \cite{Ceresole:2006iq}) that the tension of these solitonic domain walls have the correct scaling to describe a wrapped  five-brane. In all of these scenarios, the four-dimensional vacuum energy can be different on either side of the brane.  

As in the Einstein--Maxwell case, we are mainly interested in the effect of the tension term on the dynamics of the volume modulus. The appropriate source action for an $NS5$-brane wrapped on a 3-cycle of period $\Pi_3$ (we define the periods in terms of the metric $g^{(6)}$) in the four-dimensional Einstein frame is
\begin{equation}\label{eq:braneaction}
S_{sources} = M_4^3 \left( \frac{\sigma_{NS5} }{M_{10}^{6}} \right) \left( \frac{\Pi_3}{{\rm Vol(\Omega_6)}^{3/2} } \right) \int_{\Sigma} \sqrt{\gamma} \exp \left[ - \sqrt{\frac{3}{2}} \frac{\chi ({\bf \xi})}{M_4} \right] d^3 \xi.
\end{equation}
The tension of the five-branes in the original $10$ dimensional action is~\cite{Polchinski:1998rr}:
\begin{equation}\label{eq:ns5tension}
\sigma_{NS5} = \frac{(8 \pi)^{3/4}}{(2 \pi)^5} g_s^{1/2} M_{10}^6.
\end{equation} 
For $D5$-branes, the tension is obtained by replacing $g_s^{1/2}$ by $g_s^{-1}$ in the above formula. In all cases, we require that the proper volume of the 3-cycles $e^{- 3 A_0} \Pi_3 \gg M_{10}^{-3}$.
 In a strongly warped region, we can have $\Pi_3 \ll 1$ while still satisfying this constraint.

From Eq.~\ref{eq:braneaction}, we see that the tension of the dimensionally reduced brane couples to the volume modulus. Thus, just as in the toy model of Sec.~\ref{sec:toymodel}, $\chi$ picks up a spatial derivative at the location of the brane. Assuming that $\chi$ has a fixed value $\chi^*$ at the wall, the change in the spatial derivative of $\chi$ is 
\begin{equation}\label{eq:deltachiIIB}
\Delta \chi' = - M_4^2 \sqrt{\frac{3}{2}} \left( \frac{\sigma_{NS5}}{M_{10}^{5}} \right)  \left( \frac{\Pi_3}{{\rm Vol (\Omega_6)}^{3/2} } \right) \exp \left[ - \sqrt{\frac{3}{2}} \frac{ \chi^* }{M_4} \right]. 
\end{equation}

As in Section~\ref{sec:DWanalytic}, we check if the domain walls are stable by comparing this `kick' velocity to the rolling velocity at the location of the brane. We first investigate the stability of domain walls before the potential is uplifted, i.e. for domain walls separating the supersymmetric AdS vacua of the KKLT model or the non-supersymmetric AdS vacua of the BBCQ model. For both models, the potentials are physical realizations of the toy model `AdS' potential of Eq.~\ref{eq:potential}, and static domain walls can always be found. In the KKLT model, these planar domain walls preserve some supersymmetry, and correspond to the well-known BPS solutions discussed in e.g. \cite{Ceresole:2006iq}. In the BBCQ model, there is no supersymmetry guaranteeing the stability of the solutions, but nevertheless, by the arguments presented in Sec.~\ref{sec:DWanalytic} we find that the domain walls are at least stable against decompactification.

Proceeding to domain walls in the Uplifted potentials, the relevant test for a solution is again to compare the rolling velocity to the local maximum of the potential, and (in the case of planar walls) the asymptotic region $\chi \to \infty$. Starting with the latter test, we recall that stable domain wall solutions are always possible if the asymptotic falloff of the kick velocity (Eq.~\ref{eq:deltachiIIB}) is faster than that of the rolling velocity. If energy is conserved, and the minimum has zero energy, then the rolling velocity is given by  $\vert V_{uplift}\vert$ for both the KKLT and the BBCQ models. Consequently, for either the anti-$D3$- or $D7$-branes, the falloff of the rolling velocity is too fast to allow for stable planar domain walls. 

The more appropriate test is therefore to compare the kick to the rolling velocity at the location of the potential maximum, which will uniquely determine if walls are stable. The local maxima of the KKLT and the BBCQ potentials are different, and we analyze the models separately. In both cases, we approximate the height of the potential barrier with the depth of the AdS minimum, since this will approximately set the necessary size of the positive uplifting term.

The stability of domain walls in the KKLT model was investigated in Ref.~\cite{Johnson:2008vn}. Using Eq.~\ref{eq:reduceddomainaction}, we reproduce the condition for stability
\begin{equation}\label{eq:energyratioKKLT}
\frac{(\Delta \chi')}{2 \sqrt{2 |V_{\rm max}|} } \simeq \frac{\sigma_{NS5}}{M_{10}^{5}}  \frac{\Pi_3}{{\rm Vol (\Omega_6)}^{3/2} }  \frac{3}{\sqrt{2} a A} \frac{e^{ac_{\rm min}}}{c_{\rm min}} < 1.
\end{equation}
For the KKLT construction to be valid, we must have $c_{\rm min} \gg 1$. In fact, there is a relatively narrow window in which all of the approximations made in this construction are satisfied, estimated by Ref.~\cite{Freivogel:2008wm} to be roughly $10^3 < c_{\rm min} < 10^5$. At such large volume, to satisfy the criterion for stability, some of the other parameters in Eq.~\ref{eq:energyratioKKLT} must be exponentially small. From Eq.~\ref{eq:ns5tension}, the brane tension decreases at small string coupling, but obtaining an exponentially small string coupling would seem unnaturally fine-tuned (and perhaps in conflict with phenomenological constraints on the string scale). The constant $a$, if it arises from gaugino condensation on $D7$ branes where $a = 2 \pi / N_{D7}$, cannot be arbitrarily small since $N_{D7}$ cannot be large without large backreaction. In some cases, $A$ might be large~\cite{Denef:2005mm}, but the only truly viable candidate is $\Pi_3$, which in a region of strong warping scales like $\Pi_3 \sim e^{3 A_0}$.

Therefore, the stability of branes in this situation relies crucially on the period of the cycle that they wrap. If the 3-cycle is in the same throat as the anti-$D3$ branes used to uplift the potential for the volume modulus, we can estimate the period. From Eq.~\ref{eq:D3action} and ~\ref{eq:vuplift}, we see that the uplifting term scales like $e^{4A_0}$. Comparing with the depth of the AdS minimum Eq.~\ref{eq:KKLTminheight}, we must have
\begin{equation}
e^{4A_0} \propto c_{\rm min} e^{-2 a c_{\rm min}}.
\end{equation}
Therefore, the period of the three-cycle scales like
\begin{equation}
\Pi_3 \propto c_{\rm min}^{3/4} e^{-3 a c_{\rm min} / 2},
\end{equation}
which, inserting into Eq.~\ref{eq:energyratioKKLT}, is sufficiently small to guarantee stable domain walls at large volume. In conclusion, stable domain walls can be constructed only if the cycle wrapped by the brane is sufficiently warped. This result was already suggested in~\cite{Johnson:2008vn}, where a similar result for solitonic domain walls was obtained. It is also interesting to compare our result to the near-BPS domain walls in Ref.~\cite{Ceresole:2006iq}, where supersymmetry-breaking terms have to be small. This is guaranteed by a large warping in the KKLT model, which is the same condition needed for the domain wall to be stable against decompactification in our model.

Similarly, the condition for stability in the BBCQ model is
\begin{equation}\label{eq:energyratioBBCQ}
\frac{(\Delta \chi')}{\sqrt{ 2 |V_{\rm max}|}} \simeq  \frac{\sigma_{NS5}}{M_{10}^{5}}  \frac{\Pi_3}{{\rm Vol (\Omega_6)}^{3/2} }  \frac{M_4^{1/2} e^{-K_{cs} / 2}}{|A_s W_0|^{1/2}} \frac{e^{\sqrt{\frac{3}{8}} \frac{\chi_{\rm min}}{ M_4}} }{\chi_{\rm min}^{1/2} }  < 1.
\end{equation}
Here, the volume is exponentially large, e.g. in the example of ~\cite{Balasubramanian:2005zx} we have $ \chi_{\rm min} \sim 20$ (yielding a volume of about $10^{10}$ in units of $M_{10}$). Thus, to obtain stable domain walls, a significant fine-tuning of the parameters in Eq.~\ref{eq:energyratioBBCQ} is needed. Just as for the KKLT example, most of these parameters are constrained. The expectation value of the superpotential, $|W_0|$ is determined by the fluxes. $|W_0|$ need not be small in BBCQ compactifications, but can in general not be arbitrarily large, since that would require large fluxes which might violate the tadpole conditions of the compactification. Therefore, we will assume that $|W_0|$ is generically of order one, keeping in mind that larger values can be allowed for specific models. The complex structure K\"ahler potential $K_{cs}$ is determined by the unwarped periods, and is also of order one. As above, we find that the tuneable parameters are $A_s$ and  $\Pi_3$, and to make $\Pi_3$ small we need to put the corresponding cycle in a strongly warped region. Thus, even though large warping may not be necessary to construct vacua in the BBCQ model, it could be used to tune the period of the cycle wrapped by the five-brane. This seems necessary for stable domain walls. 

Before we end the discussion on the stability of domain walls in the BBCQ model, there are two things that we need to investigate. First, the $\alpha'$ correction to the K\"ahler potential couples the volume modulus to the axio-dilaton, which is a flux-fixed field. To disentangle the equations of motion for the two fields we assume that the axio-dilaton is fixed during the evolution of the volume modulus. This is motivated by the hierarchically small contribution of the $\alpha'$ corrections compared to the flux-induced potential. We leave a more careful study of these effects for future work.

Second, in order to construct a vacuum at exponentially large volume, it is crucial that the compactification manifold has at least two K\"ahler moduli~\cite{Balasubramanian:2005zx}. In the analysis above, we only considered the dynamics of the volume modulus, since this is what couples to the flux-fixed fields and the branes. However, the BBCQ potential, and also the non-trivial metric on the K\"ahler moduli space, implies that there are couplings between the volume modulus and the other K\"ahler moduli. We must therefore check, by looking at the equations of motion, that it is consistent to keep the other K\"ahler moduli fixed while the volume modulus is rolling. Performing this analysis, we find that this is possible, as long as the volume is exponentially large during the full evolution of the volume modulus. We therefore restrict our analysis to this regime, and hope to come back to the possible interplay between the K\"ahler moduli in future work.

\section{Conclusions and outlook}\label{sec:conclusions}

The dynamics of the volume modulus can have important implications for the picture of eternal inflation in the string theory landscape. In this paper, we have shown that only a rather restricted set of configurations of the internal manifold and sources can be connected across stable planar or spherical domain walls.

The dilatonic coupling between the volume modulus and domain walls separating different four-dimensional vacua can completely overwhelm the stabilizing potential in the neighborhood of the domain wall. Using analytic methods, we have shown that non-singular static planar domain wall and SO(3,1) bubble solutions exist only when the energy scale associated with the tension of the domain wall is smaller than the energy scale associated with the potential maximum for the volume modulus. Rephrasing this condition, solutions do not exist when the tension of domain walls separating compactified vacua is larger than the tension of domain walls interpolating between the compactified and decompactified phases, indicating that it is energetically favorable for such domain walls to decompactify.

Using a set of numerical simulations, we find that this is indeed the case. When non-singular solutions do not exist, domain wall and bubble solutions seed a growing region of the decompactified phase. In the case of bubble solutions, when the true vacuum has zero or positive energy, the entire bubble interior goes over to the decompactified phase. This completely removes the spacetime region containing the true vacuum. Regardless of the transition mechanism that produces the original region of true vacuum, the true vacuum phase will not exist at late times, and therefore cannot be populated by any such transition.

In a spacetime without a cosmological constant, the presence of an unstable domain wall separating different four-dimensional vacua causes all of space to decompactify. If there is a cosmological constant, roughly a comoving Hubble volume of the background space will be converted to the decompactified phase at late times. In neither case does the spacetime evolve to a solution where different four-dimensional vacua coexist. This conclusion is independent of the creation mechanism, be it the formation of domain walls via the Kibble mechanism, or the nucleation of bubbles of true vacuum during eternal inflation.

On the other hand, our analysis shows that domain walls separating AdS vacua are not troubled by this instability. Although we have not used it in our analysis, it is interesting to note that such domain walls can be BPS, in which case their stability is guaranteed by supersymmetry. Our analysis is sufficient to show that such domain walls are not unstable to decompactification, but there may exist instabilities in other moduli fields if they couple to the domain walls. 

The dilatonic coupling of the volume modulus can also play an important role in determining the outcome of a collision between two bubbles. Such collisions might leave detectable experimental signatures of eternal inflation, but only if our observed cosmology can exist to their future. For stable bubble walls, the energy released in a collision has the ability to seed a region where space decompactifies. We have performed a number of numerical simulations, which indicate that this process is in fact quite efficient. Depending on the kinematics and properties of the potential, the decompactified region can accelerate into the bubble interior, rendering such collisions unobservable. Further analysis of these preliminary results will undoubtedly be an important element in determining the observability of bubble collisions arising in context of eternal inflation driven by the string theory landscape.  

We have investigated compactifications of Einstein--Maxwell and type IIB string theory, which both give rise to landscapes of vacua, domain walls, and possibly eternal inflation. Fundamental branes act as domain walls separating four-dimensional vacua in these theories, and possess a  dilatonic coupling to the volume modulus. We find that domain walls between dS or Minkowski vacua can be unstable to decompactification, as described above. 

The dilatonic coupling becomes especially problematic when the potential fixing the volume modulus is small compared to the tension of the walls. This is a common feature for both KKLT and BBCQ compactifications of type IIB string theory since the volume modulus is unfixed at leading order. In this case, we find that stable domain walls and bubbles arise only when the brane wraps a strongly warped cycle, leading to a reduced tension. The instabilities provide further motivation for finding dS vacua in compactifications where all geometric moduli are fixed at the same scale, as is the case in e.g. generic type IIA compactifications. These compactifications bear some resemblance to the Einstein--Maxwell model, where we find that stability depends on the dimensionality of the cycles that a fundamental brane wraps in the internal manifold.  

The existence of unstable domain walls separating different four-dimensional vacua leads us to conclude that only a restricted set of configurations of the internal manifold and sources can be populated by eternal inflation. However, our results do not exclude the existence of solutions where vacua of different dimensionality, separated by domain walls, coexist. Such a solution could contain four-dimensional vacua, separated by intervening regions where a different number of dimensions is kept large, as suggested in Ref.~\cite{Carroll:2009dn}. Further exploration of these and other ideas will be necessary to determine the full dynamical role of extra dimensions in populating landscapes of vacua.

\begin{acknowledgments}
The authors wish to thank J. Blanco-Pillado, B. Freivogel, J. Garriga,   U. Yajnik for helpful conversations. AA's research is funded by NSF grant PHY-0757912, and a ``Foundational Questions in Physics and CosmologyÕÕ grant from the Templeton Foundation. MJ is supported by the Gordon and Betty Moore Foundation, and thanks the Perimeter Institute for their hospitality while portions of this work were completed. The research of ML  is supported by the Deutsche Forschungsgemeinschaft.
\end{acknowledgments}

%%%%%%%%%%%%%%%%%%%
\bibliography{runawaydomainwalls}

\end{document}